\documentclass[twocolumn]{IEEEtran}
\usepackage{cite}
\usepackage{amsmath, amssymb, amsfonts, amsthm}
\usepackage{algorithm}
\usepackage[noend]{algorithmic}
\usepackage{graphicx}
\usepackage{textcomp}
\usepackage{xcolor}
\usepackage{color}
\usepackage{bm}
\usepackage{subfigure} 
\usepackage{booktabs}
\usepackage{autobreak}
\usepackage{multirow}
\usepackage{booktabs}
\usepackage{authblk}
\usepackage{mathrsfs}
\usepackage{array}
\usepackage{makecell} 
\usepackage{float}
\usepackage{soul}
\usepackage{stmaryrd}
\usepackage{dsfont}
\usepackage{bbm}
\usepackage[hidelinks]{hyperref}
\usepackage{cleveref}
\usepackage{pifont}
\usepackage{enumitem}
\usepackage{times}

\newtheorem{lemma}{Lemma}

\newtheorem{theorem}{Theorem}

\definecolor{deepgreen}{rgb}{0.0, 0.5, 0.0}

\definecolor{blue}{rgb}{0.0, 0.0, 0.0}

\begin{document}
\title{\fontsize{22.5pt}{26pt}\selectfont Conformal Distributed Remote Inference in Sensor Networks Under Reliability and Communication Constraints}
\author{
Meiyi Zhu, Matteo Zecchin \IEEEmembership{Member, IEEE}, Sangwoo Park \IEEEmembership{Member, IEEE}, Caili Guo \IEEEmembership{Senior Member, IEEE}, Chunyan Feng \IEEEmembership{Senior Member, IEEE}, Petar Popovski \IEEEmembership{Fellow, IEEE}, Osvaldo Simeone \IEEEmembership{Fellow, IEEE}

\vspace{-0.5cm}

\thanks{
The work of M. Zhu, C. Guo and C. Feng was supported by the National Natural Science Foundation of China (62371070), by the Beijing Natural Science Foundation (L222043), by the National Natural Science Foundation of China (61871047), and by the BUPT Excellent Ph.D. Students Foundation (CX2023150).
The work of M. Zecchin and O. Simeone was supported by the European Union's Horizon Europe project CENTRIC (101096379). The work of Petar Popovski was supported in part by the Villum Investigator Grant ``WATER'' from the Velux Foundations, Denmark. The work of O. Simeone was also supported by an Open Fellowship of the EPSRC (EP/W024101/1), and by the EPSRC project (EP/X011852/1).

Meiyi Zhu, Caili Guo and Chunyan Feng are with the Beijing Key Laboratory of Network System Architecture and Convergence, School of Information and Communication Engineering, Beijing University of Posts and Telecommunications, Beijing 100876, China (e-mail: lia@bupt.edu.cn; guocaili@bupt.edu.cn; cyfeng@bupt.edu.cn).

Matteo Zecchin, Sangwoo Park, and Osvaldo Simeone are with the King's Communications, Learning \& Information Processing (KCLIP) lab within the Centre for Intelligent Information Processing Systems (CIIPS), Department of Engineering, King's College London, London WC2R 2LS, U.K. (e-mail: \{matteo.1.zecchin, sangwoo.park, osvaldo.simeone\}@kcl.ac.uk).

Petar Popovski is with the Connectivity Section, Aalborg University, Aalborg, Denmark. (e-mail: petarp@es.aau.dk)
}
}

\maketitle

\begin{abstract}
This paper presents communication-constrained distributed conformal risk control (CD-CRC) framework, a novel decision-making framework for sensor networks under communication constraints. Targeting multi-label classification problems, such as segmentation, CD-CRC dynamically adjusts local and global thresholds used to identify significant labels with the goal of ensuring a target false negative rate (FNR), while adhering to communication capacity limits. CD-CRC builds on online exponentiated gradient descent to estimate the relative quality of the observations of different sensors, and on online conformal risk control (CRC) as a mechanism to control local and global thresholds. CD-CRC is proved to offer deterministic worst-case performance guarantees in terms of FNR and communication overhead, while the regret performance in terms of false positive rate (FPR) is characterized as a function of the key hyperparameters. Simulation results highlight the effectiveness of CD-CRC, particularly in communication resource-constrained environments, making it a valuable tool for enhancing the performance and reliability of distributed sensor networks.
\end{abstract}


\section{Introduction}
\begin{figure*}[htbp]
    \centering
    {\includegraphics[width = 0.75\textwidth]{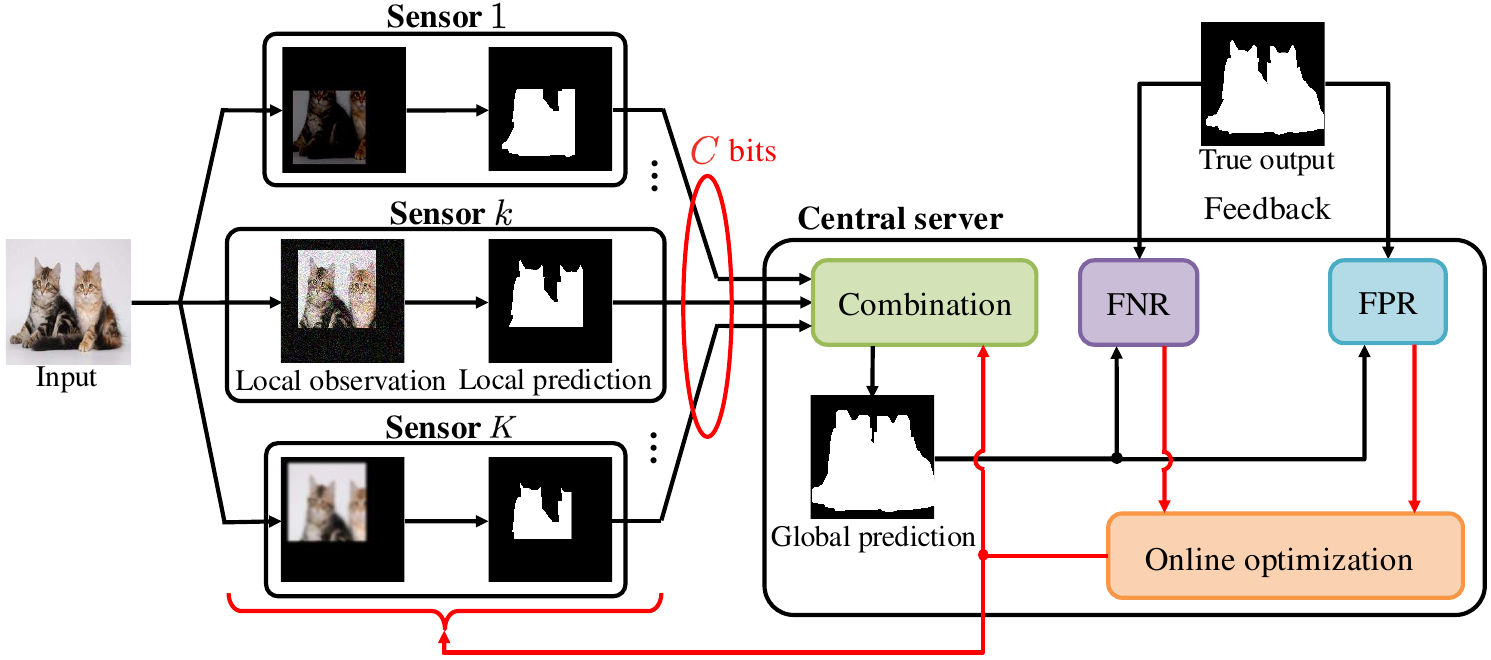}}
    \caption{Illustration of the system model and design problem: Each sensor generates a local prediction for a general multi-label problem, e.g., segmentation, based on a partial and noisy observation of the underlying input. The local predictions are transmitted through a shared channel to a central server, which combines them to obtain a global prediction. After the decision is made, feedback on the correct decision makes it possible to compute FNR and FPR at the central server. The system is optimized online to minimize long-term FPR, while meeting long-term constraints on FNR and channel capacity.}\label{fig_sys_mod}
\end{figure*}
Distributed sensor networks are a cornerstone of modern Internet-of-Things (IoT) systems, enabling efficient information processing through geographically dispersed sensors \cite{behera2019sep, muhammad2018secure}. Advanced sensors, such as cameras, have expanded their functionality across diverse applications, including user localization for beam management in wireless systems \cite{sun2019energy, gong2019optimal}, obstacle detection in vehicular networks \cite{hansen2017driver, tedeschini2023cooperative}, and intelligent traffic infrastructure \cite{kamtue2024phyot, wang2021rodnet, mao2021moving}. In these scenarios, sensors capture various environmental aspects, and a central server aggregates their inputs for decision-making.

\vspace{-3mm}
\subsection{Distributed Remote Inference}
We focus on a scenario in which multiple sensors observe different features of the underlying input. For example, vision sensors may have different partial views of the scene of interest, and they may be subject to various random impairments, such as darkness, noise artifacts, and blurring. {\color{blue}The goal is to perform \textit{multi-label classification} tasks--where each instance can be associated with multiple labels simultaneously, such as binary segmentation--through communication to a central server.}

In binary segmentation, as shown in Fig. \ref{fig_sys_mod}, one wishes to produce a binary mask that distinguishes objects of interest and background. In the system under study, based on the respective local observations, each sensor produces a local binary mask. Based on limited communication from sensors to server, the central server fuses information about the local predictions to generate a global binary mask that estimates the space covered by the objects of interest.

In this context, most existing work leverages conventional sensor fusion methods, such as weighted averaging, Bayesian inference, or Kalman filtering, to aggregate local predictions from multiple sensors \cite{hall1997introduction, rammelkamp2020low, wang2021multi, gostar2020centralized}. The fusion process usually involves assigning weights to each sensor's output based on the sensor's perceived reliability, with the goal of optimizing the overall prediction performance \cite{katenka2007local, niu2006fusion, sharma2020sensor, ravindran2022camera}. In some cases, machine learning-based approaches, such as deep learning or ensemble methods, are employed to enhance fusion accuracy by learning patterns from historical data \cite{su2015multi, liu2024semantic, lan2022progressive}.

However, these existing approaches typically assume statistical models for the underlying source signal and for the sensor observations, and focus on optimizing statistical criteria such as expected error rates or mean square error. For instance, \cite{gostar2020centralized} utilizes labeled multi-Bernoulli filters based on Bayesian principles, incorporating prior distributions for target existence and dynamics, while \cite{ravindran2022camera} leverages prior knowledge in a Bayesian neural network framework to improve sensor fusion accuracy. Accordingly, unless one makes strong modeling assumptions, none of the existing schemes provide formal performance guarantees. In contrast, this paper focuses on providing deterministic \textit{worst-case} guarantees that are \textit{assumption-free} and do not rely on statistical models.

\subsection{Distributed Conformal Risk Control}
To ensure deterministic, assumption-free reliability, distributed conformal prediction (D-CP) was introduced in \cite{gasparin2024conformal} for settings without any communication constraints between sensors and central server. Focusing on conventional single-label classification problems, D-CP leverages online conformal prediction (CP) \cite{gibbs2021adaptive} to dynamically adjust a shared local decision threshold used by all the sensors to construct a prediction set. D-CP ensures the deterministic coverage requirement that the true label is included in the prediction set with a target long-term rate. Furthermore, D-CP employs an online exponentiated gradient strategy \cite{de2014follow} to sequentially adjust weights used by the central server to combine the sensors' decisions, aiming to minimize the size of the prediction set.

In this paper, we introduce a novel strategy that extends the scope of D-CP by accounting for the presence of communication constraints between sensors and central server. The proposed method, referred to as \emph{communication-constrained distributed conformal risk control} (CD-CRC), targets a general multi-label classification problem, providing assumption-free long-term deterministic guarantees on false negative rate (FNR) and communication load.

Due to the presence of communication constraints, CD-CRC differs from D-CP in several fundamental aspects, including the use of multiple local thresholds and the simultaneous update of a global threshold. These novel aspects pose technical challenges both in terms of the design of the online optimization strategies applied at sensors and central server, and in terms of the performance analysis.

\subsection{Related Work}
Here, we briefly review additional relevant papers related to the theme of this work.

\textit{Sensor Fusion:}
Sensor fusion methods are typically categorized into data fusion, feature fusion, and decision fusion, depending on the type of information transmitted to the central server.

\textit{Data fusion} combines raw data from multiple sensors before processing at the central server \cite{hall1997introduction}. Techniques include low-level fusion, which combines raw sensor data using methods like Kalman filtering for noise reduction and signal enhancement \cite{rammelkamp2020low}; track-to-track fusion, which aggregates intermediate results, such as sensor-generated tracks, to improve tracking accuracy and reduce false alarms \cite{wang2021multi}; and random finite set based fusion, which is particularly effective in multi-object tracking under variable target counts and limited sensor views \cite{gostar2020centralized}. While accurate, data fusion incurs high communication costs.

\textit{Feature fusion} integrates features extracted from sensors to form a joint representation, reducing the dimensionality of transmitted data, while preserving complementary information. For example, reference \cite{su2015multi} proposes multi-view convolutional neural networks to combine features for 3D object recognition, and the more recent work focuses on balancing communication efficiency and inference accuracy through information bottleneck-based compression \cite{liu2024semantic} and progressive feature transmission \cite{lan2022progressive}.

\textit{Decision fusion} aggregates decisions made independently by each sensor and combines them at a central server. This class of methods is robust to sensor failures and noise, with key works focusing on improving accuracy using techniques like weighted majority voting \cite{katenka2007local}, likelihood ratio testing \cite{niu2006fusion, sharma2020sensor}, and Bayesian decision theory \cite{ravindran2022camera}. These approaches aim to minimize overall error while maintaining low communication costs.

As anticipated, these existing methods provide performance guarantees in terms of statistical averages, only under strong probabilistic modeling assumptions. In contrast, this paper focuses on providing assumption-free, worst-case performance guarantees that ensure long-term reliability, without requiring the validity of statistical models.

\textit{Conformal Risk Control:} {\color{blue}CP is a statistical framework that provides valid measures of predictive uncertainty by generating prediction intervals or sets that contain the true label with a specific probability \cite{vovk2005algorithmic, barber2021predictive}.} CP has since been extended to distributed settings, particularly in federated scenarios, where data remains private across multiple devices while reliable decision-making is performed at the central server \cite{FedCPQQ, zhu2024federated}. These approaches operate offline and rely on data exchangeability. Online CP, developed in \cite{gibbs2021adaptive} and further studied in \cite{bhatnagar2023improved, angelopoulos2024online, gibbs2024conformal}, maintains long-term calibration without statistical assumptions on data generation. This way, online CP ensures the desired coverage frequency over time, even under adversarial data distribution shifts.

While CP focuses on maintaining coverage, conformal risk control (CRC) can address more general risk metrics, allowing control of risks such as FNR at any specified level \cite{angelopoulos2022conformal, cohen2024cross, zecchin2025generalization}. Online versions of CRC were introduced in \cite{feldman2022achieving, angelopoulos2024conformal}.

\subsection{Contributions and Organization}
In this paper, we introduce CD-CRC, a novel distributed multi-label classification strategy that ensures long-term performance guarantees in terms of FNR, while maintaining FPR under communication constraints. The main contributions of this paper are summarized as follows.
\begin{itemize}
    \item We adapt the state-of-the-art distributed conformal inference scheme, D-CP \cite{gasparin2024conformal}, to the multi-label problem under study. This yields the benchmark, distributed conformal risk control (D-CRC) protocol, for settings without communication constraints. D-CRC adopts a shared local threshold, generalizing online CRC \cite{feldman2022achieving} to a distributed setup. We provide a theoretical analysis of D-CRC, including guarantees for the FNR and bounds on the FPR, which are tailored from \cite{gasparin2024conformal} to the multi-label setting.
    \item We propose CD-CRC, a novel protocol that dynamically updates local and global thresholds along with combining weights at the central server to satisfy both FNR and capacity constraints, while minimizing FPR.
    \item We provide an analysis of the performance of CD-CRC, providing deterministic, assumption-free guarantees for both the FNR and the communication load, and establishing an upper bound on the FPR.
    \item Simulation results demonstrate the advantage of CD-CRC over existing methods, particularly in scenarios with constrained communication resources.
\end{itemize}

The remainder of this paper is organized as follows. Sec. \ref{sec_sys_prob} outlines the setting and defines the problem. Sec. \ref{sec_D_CRC} presents the D-CRC scheme, which extends D-CP in \cite{gasparin2024conformal} to multi-label classification, operating under no communication constraints, and analyzes its performance. In Sec. \ref{sec_CD_CRC}, we propose CD-CRC, detailing design guidelines and providing performance guarantees. Sec. \ref{sec_experiments} presents experimental results comparing the performance of CD-CRC with benchmarks, demonstrating its effectiveness. Finally, Sec. \ref{sec_conclusion} concludes the paper and suggests directions for future work.

\section{System Model and Problem Definition} \label{sec_sys_prob}
\subsection{Setting}
\begin{figure*}[htbp]
    \centering 
    {\includegraphics[width = 0.8\textwidth]{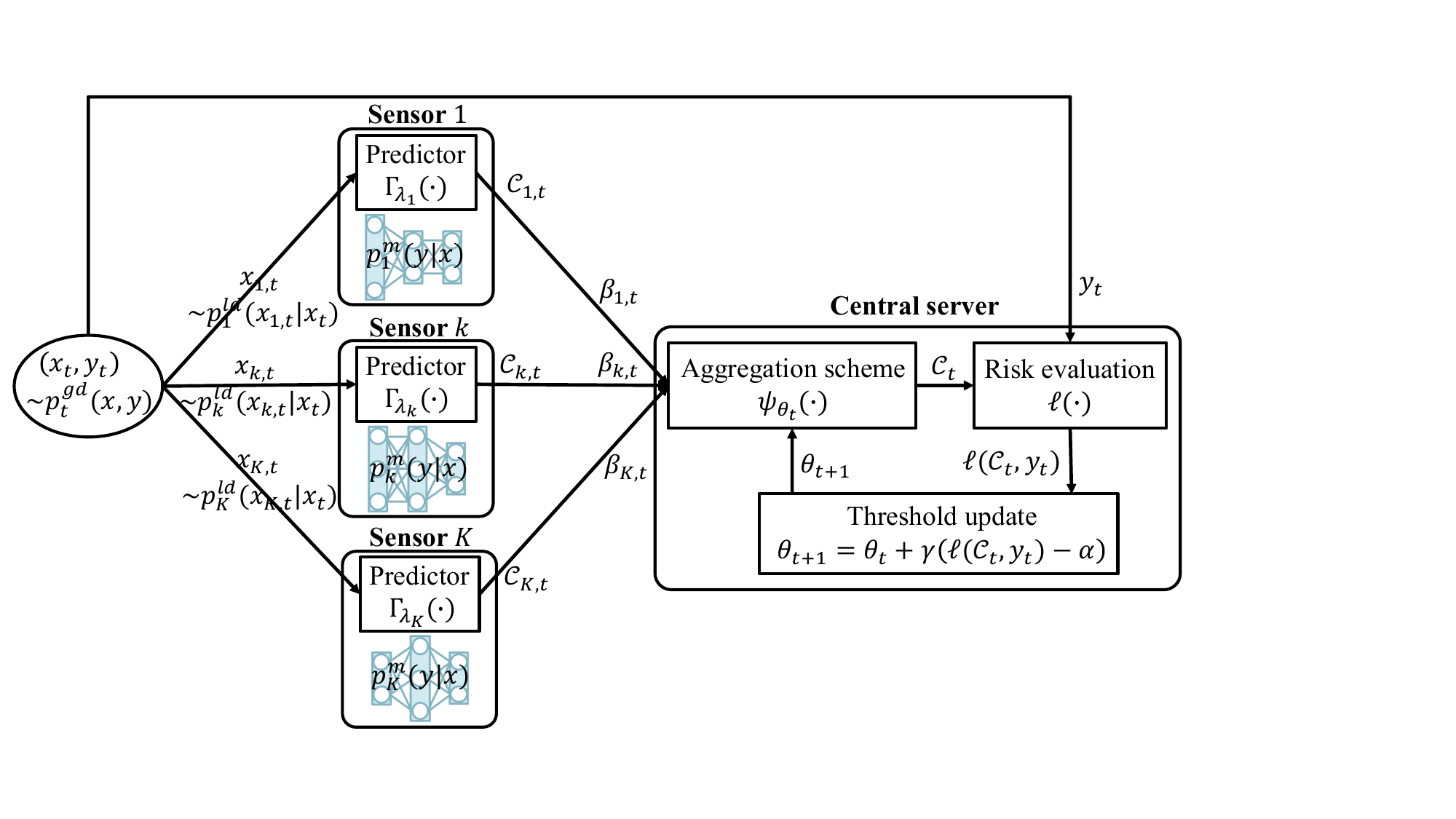}}
    \caption{Illustration of the overall workflow of the studied distributed multi-label classification system.}\label{fig_workflow}
\end{figure*}
As depicted in Fig. \ref{fig_workflow}, we consider a sensor network consisting of $K$ sensors and a central server, whose goal is to carry out distributed multi-label prediction under reliability and communication constraints. Specifically, at each discrete time $t$, the system operates as follows:
\begin{itemize}
    \item[1.] The sensors send information to the central server.
    \item[2.] The central server addresses a multi-label classification problem, such as segmentation. Depending on the application, the server may act on the classification decision. For instance, in a wireless communication setting, the central server may use the segmentation decision to direct the antenna beams towards the mobile users \cite{feng2021uav}.
    \item[3.] The central server obtains feedback on the false negative rate and false positive rate performance of its classification decision, which will be formally defined below. This feedback may be obtained as a result of the actions taken based on the classification decisions. For example, in the mentioned wireless use case, the server may be informed about the fraction of users that did not receive timely service.
\end{itemize}

In more detail, at each discrete time $t$, each sensor $k$ collects an input $X^t_k$ corresponding to a different noisy view of an underlying input $X^t$, which is associated with a label vector $Y^t\in\left\{0,1\right\}^L$. The goal is to use limited communication from sensors to the central server to support reliable detection of vector $Y^t$ at the server.

This setting captures multi-label classification tasks, in which there are $L$ possible labels describing the scene, e.g., $L$ possible objects, and the target variable $Y^t$ indicates which labels are pertinent to $X^t$. As shown in Fig. \ref{fig_sys_mod} and Fig. \ref{fig_workflow}, this formulation encompasses the segmentation of an image, in which each image $X^t$ with $L$ pixels is associated with a binary mask $Y^t\in\left\{0,1\right\}^L$ specifying the position of an object of interest. In particular, an element $Y^t_l = 1$ indicates that the pixel $l$ is part of the target, whereas $Y^t_l=0$ identifies the pixel $l$ as background. Henceforth, we will refer to each entry $Y^t_l$ of the vector $Y^t$ as a label.

Each sensor $k$ is equipped with a pre-trained machine learning model $g_k(\cdot)$ that generates a local inference decision based on the local observation $X_k^t$. For each time step $t$, based on the observation $X^t_k$ of the target object $X^t$, the local model at sensor $k$ outputs the vector of probabilities
\begin{align}\label{eq_loc_prob_vec}
    S^t_k = g_k(X_k^t)
    \in[0,1]^L,
\end{align}
where each $l$-th entry $S^t_{k,l}\in[0,1]$ represents the estimated probability that label $l$ is pertinent for input $X^t$, i.e., that we have $Y^t_l=1$.

In order to facilitate communication of the local decision $S^t_k$ to the server, based on a local threshold $\lambda_k^t$, each sensor $k$ computes the binary prediction vector
\begin{align}\label{eq_loc_pred_vec}
    U^t_k=\mathds{1}\left\{S_k^t\geq \lambda_k^t\right\} \in \left\{0,1\right\}^L,
\end{align}
where the indicator function $\mathds{1}\left\{\cdot\right\}$ returns $1$ if the augment is true and $0$ otherwise, and the inequality in \eqref{eq_loc_pred_vec} is applied element-wise. Vector $U^t_k$ contains hard decisions made by sensor $k$ about the relevance of each label for the current input $X^t$. The vector $U^t_k$ is then transmitted to the server.

In order to aggregate the local predictions from all $K$ sensors, the server utilizes a set of weights $\left\{\beta_k^t\right\}_{k=1}^K$, where $\beta^t_k\in[0,1]$ and $\sum_{k=1}^K\beta_k^t=1$, to obtain the weighted decision vector
\begin{align}\label{eq_glo_prob_vec}
    R^t = \sum_{k=1}^K \beta^t_k U^t_k\in[0,1]^L.
\end{align}
The weight $\beta^t_k$ reflects the contribution of the sensor $k$ to the global decision \eqref{eq_glo_prob_vec}, and thus it should be ideally designed depending on the quality of the observation {\color{blue}$X^t_k$} at sensor $k$. Finally, based on the soft estimate \eqref{eq_glo_prob_vec}, using a global threshold $\theta^t$, the server obtains hard decisions for each label as
\begin{align}\label{eq_glo_pred_vec}
    V^t = \mathds{1}\left\{R^t\geq \theta^t\right\}\in\left\{0,1\right\}^L,
\end{align}
where the indicator function $\mathds{1}\left\{\cdot\right\}$ returns $1$ if the argument is true and $0$ otherwise. Entry $V^t_l=1$ indicates that label $l$ is estimated as being relevant for input $X^t$, while $V^t_l=0$ signifies that label $l$ is identified as irrelevant for the current input.

\subsection{Design Problem}
The goal of this work is to develop online optimization algorithms for the update of: \textit{(i)} the local thresholds $\left\{\lambda^t_k\right\}_{k=1}^K$, \textit{(ii)} the combining weights $\left\{\beta_k^t\right\}_{k=1}^K$, and \textit{(iii)} the global threshold $\theta^t$. We aim at minimizing the FPR, i.e., the fraction of incorrectly detected labels, while imposing constraints on the FNR, i.e., on the fraction of missed labels, as well as the communication overhead. We take a worst-case design approach, making no assumption on the joint distribution of source $X^t$, observations $\left\{X_k^t\right\}_{k=1}^K$, and multi-label vector $Y^t$.

\subsubsection{False negative rate and false positive rate} To justify the problem design formulation, consider again the prediction problem in Fig. \ref{fig_sys_mod}. For this application, the proposed methodology imposes a constraint on the fraction of target pixels that the central server's prediction allows to miss, i.e., the FNR, while attempting to maximize the fraction of target pixels that are correctly predicted.

In general, the FNR measures the proportion of labels erroneously classified as irrelevant, i.e., {\color{blue}$V^t_l=0$}, while they are, in fact, pertinent, i.e., {\color{blue}$Y^t_l=1$ for label $l$}. As mentioned, in the segmentation example of Fig. \ref{fig_workflow}, the FNR is the fraction of missed pixels containing the object of interest. The FNR is calculated as
\begin{align}\label{eq_FNR}
    N^t=\frac{\sum_{l=1}^L Y^t_l\left(1-V_l^t\right)}{\sum_{l=1}^LY^t_l},
\end{align}
where the numerator counts the number of labels erroneously predicted as irrelevant, and the denominator counts the number of ground-truth pertinent labels.
Note that a global prediction $V^t$ consisting entirely of ones yields the minimal FNR of $0$, indicating no false negatives; however, it clearly lacks discriminative power.

To measure the discriminative power of a protocol, the FPR evaluates the fraction
\begin{align}\label{eq_FPR}
    P^t = \frac{\sum_{l=1}^L\left(1-Y^t_l\right)V^t_l}{\sum_{l=1}^L\left(1-Y^t_l\right)},
\end{align}
where the numerator counts the number of labels incorrectly classified as pertinent, and the denominator counts the number of ground-truth irrelevant labels. A trivial prediction $V^t$ with all ones would return an FPR equal to $1$.

\subsubsection{Communication constraint}\label{subsec_com_cons}
Each $L\times 1$ binary prediction vector $U^t_k$ in \eqref{eq_loc_pred_vec} must be communicated from sensor $k$ to the server. If no compression strategy is implemented, this requires the transmission of $L$ bits from each sensor. However, when there exist communication constraints between sensors and the server, this approach is generally inefficient. In fact, without any compression, sensors that do not obtain any useful information would require the same bandwidth as sensors that have non-trivial local decisions to communicate. Given that our main design goal is ensuring a constraint on the FNR \eqref{eq_FNR}, a local decision vector $U^t_k$ is deemed to be uninformative if it contains only ones. In fact, as discussed, in this case any FNR level is guaranteed, but without any discriminative power.

Accordingly, we focus on compression strategies that require zero communicated bits when the decision corresponds to an all-one vector $U^t_k$, i.e.,
\begin{align}\label{encode_require1}
    B^t_k=0~\textrm{for}~\lambda^t_k\leq 0,
\end{align}
while the number of communicated bits $B^t_k$ increases towards its maximum value for vectors $U^t_k$ with an increasing fraction of entries equal to $0$. In other words, when the threshold increases from $0$ to $1$, the number of communicated bits increases. Normalizing the number of bits by the size of the encoded vector, the maximum value for the communication overhead $B^t_k$ is set as
\begin{align}\label{encode_require2}
    B^t_k=1~\textrm{for}~\lambda^t_k\geq 1,
\end{align}
since a threshold $\lambda^t_k\geq 1$ yields an all-zero decision $U^t_k$. Accordingly, the communication cost per sensor refers to the average number of bits per element of the decision vector. Moreover, the overall communication cost from the $K$ sensors to the server at time $t$ is equal to the sum $\sum_{k=1}^KB^t_k\in[0,K]$.

\subsubsection{Design problem} The goal of this study is to minimize the long-term FPR \eqref{eq_FPR}, under deterministic long-term constraints on FNR and communication capacity. The problem is formulated as\begin{subequations}\label{eq_obj}
    \begin{align}
        &\underset{\left\{\left\{\lambda_k^t\right\}_{k=1}^K, \left\{\beta_k^t\right\}_{k=1}^K, \theta^t\right\}_{t=1}^T}{\textrm{minimize}}\hspace{-0.2cm}\lim_{T\rightarrow\infty}\frac{1}{T}\sum_{t=1}^T P^t \label{subeq_FPR}\\
        &\text{subject to:}~\lim_{T\rightarrow \infty} \frac{1}{T} \sum_{t=1}^T N^t \leq \alpha\label{subeq_FNR}\\
        &\quad\quad\quad\text{and}~\lim_{T\rightarrow \infty} \frac{1}{T} \sum_{t=1}^T \sum_{k=1}^K B^t_k \leq C\label{subeq_capacity},
    \end{align}
\end{subequations}
where $\alpha$ is the maximum tolerated average FNR and $C\in[0,K]$ is the available long-term channel capacity. Note that with $C\geq K$, the communication constraint \eqref{subeq_capacity} can be removed, since all sensors can communicate all $L$ bits at each time $t$. In contrast, with $C<K$, the sensors need to leverage the properties of the communication scheme desired above to opportunistically reduce the local thresholds in order to meet the requirement \eqref{subeq_capacity}.

We note that, while we will focus on the FPR objective in \eqref{subeq_FPR}, the proposed methodology can be adapted directly to any objective of the form $a(Y^t) \sum_{l=1}^L b\left(Y^t_l\right)V^t_l$ with non-negative functions $a(\cdot)$ and $b(\cdot)$. Such an objective recovers FPR in \eqref{eq_FPR} by setting $a\left(Y^t\right)=1/\sum_{l=1}^L\left(1-Y^t_l\right)$ and $b\left(Y^t_l\right)=\left(1-Y^t_l\right)$, but also alternative objectives, such as the prediction set size $\sum_{l=1}^L V^t_l$, which is obtained with $a(Y^t)=b(Y_l^t)=1$ \cite{gasparin2024conformal}.

\subsubsection{Online optimization}\label{subsec_online}
We wish to address problem \eqref{eq_obj} in an online fashion by leveraging feedback or past decisions. Specifically, at each time $t$, upon making a decision $V^t$, the server receives feedback about the FNR \eqref{eq_FNR} and the FPR \eqref{eq_FPR}. Importantly, feedback is collected only after a decision is made, and only about the values of the FNR and the FPR, not about the original multi-label $Y^t$. Based on the received feedback, the optimization variables $\left\{\beta_k^t\right\}_{k=1}^K$, $\left\{\lambda_k^t\right\}_{k=1}^K$, and $\theta^t$ are updated with the goal of addressing problem \eqref{eq_obj}.

The assumed feedback framework captures many settings of practical interest. For example, in wireless communication, segmentation decisions used to adjust antenna beams toward mobile users can be assessed based on feedback from the users \cite{feng2021uav}. As another application, in security and surveillance use cases, threat detection decisions can be evaluated in hindsight using feedback from human inspection or forensic evidence \cite{rezaee2024survey}.

As a final remark, as we will further discuss in Sec. \ref{sec_theo}, meeting the FNR requirement \eqref{subeq_FNR} does not require the exact feedback of the FNR. In fact, a conservative estimate of the FNR, i.e., an upper bound on the FNR \eqref{eq_FNR}, is sufficient to ensure the reliability requirement \eqref{subeq_FNR}.

\section{Distributed Conformal Risk Control}\label{sec_D_CRC}
This section describes distributed conformal risk control (D-CRC), a direct extension of the D-CP scheme introduced in \cite{gasparin2024conformal} from single-label ($L=1$) to multi-label classification. D-CRC serves as the main benchmark in our study, addressing problem \eqref{eq_obj} for the special case in which there are no communication constraint \eqref{subeq_capacity}. We then present the theoretical performance of D-CRC by tailoring the results in \cite{gasparin2024conformal} to the analysis of the FNR and FPR performance for multi-label classification.

\subsection{Local Threshold Update} \label{sec_D_CRC_loc_thre_up}
In D-CRC, the global threshold $\theta^t$ is fixed to $\theta^t=\theta\in(0,1]$ for all time $t\geq 1$\footnote{In \cite{gasparin2024conformal}, $\theta$ is selected as $1/2$ but can be generalized to any values in $(0,1]$ according to \cite{gasparin2024merging}.}, while a common local threshold across all sensors,
\begin{align}\label{eq_loc_thre_DCRC}
    \lambda^t_1 = \cdots = \lambda^t_K = \lambda^t,
\end{align}
is optimized to meet the long-term FNR constraint in \eqref{subeq_FNR}. Specifically, the common local threshold is updated by following online CRC as \cite{angelopoulos2022conformal, feldman2022achieving}
\begin{align}\label{eq_loc_thre_up_DCRC}
    \lambda^{t+1} = \lambda^t - \rho\left(N^t-\alpha\right),
\end{align}
where $\rho>0$ represents the step size.

By \eqref{eq_loc_thre_up_DCRC}, if the FNR for the current decision $V^t$ at the server, which is given by $N^t$, exceeds the constraint $\alpha$, the local threshold $\lambda^{t+1}$ for all sensors is decreased, leading to more labels being classified as relevant in the local prediction vector $U^{t+1}_k$ in \eqref{eq_loc_pred_vec}. This update has the effect of increasing the entries of the soft estimate $P^{t+1}$ in \eqref{eq_glo_prob_vec}, resulting in more labels deemed as relevant in the global prediction vector $V^{t+1}$ in \eqref{eq_glo_pred_vec}, and reducing the global FNR, $N^{t+1}$, at the next time. Conversely, if the current FNR, $N^t$, is below the constraint, the local threshold $\lambda^{t+1}$ is increased so as to approach the FNR tolerance $\alpha$.

It is emphasized that this approach does not account for any communication constraint, making it possible to tune the local thresholds solely to meet the reliability requirement in \eqref{subeq_FNR}.

\subsection{Combining Weights Update}\label{sec_D_CRC_wei_up}
The weighted combining rule in \eqref{eq_glo_pred_vec} from $K$ sensors can be interpreted as a \textit{weighted majority voting procedure} from $K$ \textit{experts} \cite{gasparin2024conformal}. Each sensor $k$ acts as an expert, providing local decisions based on the local observation $X^t_k$ of the input $X^t$. The server aggregates local decisions to form a final, more reliable conclusion. Ideally, the server should give larger weights $\beta^t_k$ in the centralized decision \eqref{eq_glo_prob_vec} to more reliable sensors.

Based on this observation, in order to minimize the long-term FPR in \eqref{subeq_FPR}, D-CRC adopts the \textit{online exponentiated gradient rule} \cite{orabona2019modern, de2014follow} that updates the combining weights as
\begin{align}\label{eq_wei_up_DCRC}
    \beta^{t+1}_k=\frac{\exp\left\{-\eta^{t+1}\sum_{\tau=1}^tP^{\tau}_k\right\}}{\sum_{k=1}^K\exp\left\{-\eta^{t+1}\sum_{\tau=1}^tP^{\tau}_k\right\}},
\end{align}
where $\eta^t>0$ denotes an adaptive learning rate and $P^t_k$ evaluates the local FPR, defined in a similar manner to \eqref{eq_FPR} as
\begin{align}\label{eq_loc_FPR}
    P^t_k = \frac{\sum_{l=1}^L\left(1-Y^t_l\right)U^t_{k,l}}{\sum_{l=1}^L\left(1-Y^t_l\right)}.
\end{align}
The update rule \eqref{eq_wei_up_DCRC} increases the weight $\beta^{t+1}_k$ of the sensor $k$ that has achieved the lowest average FPR so far.

Following the theory of online convex optimization \cite{vovk2001competitive, cesa2006prediction}, the learning rate $\eta^t$ is updated as
\begin{align}\label{eq_lr_up_DCRC}
    \eta^{t+1}=\frac{\ln K}{\Delta^t},
\end{align}
where
\begin{align}\label{eq_Delta_DCRC}
    \Delta^t \hspace{-1mm}=\hspace{-1mm} \sum_{\tau=1}^t\hspace{-1mm}\left(\sum_{k=1}^K \beta^{\tau}_kP^{\tau}_k - \hspace{-1mm}\left(\hspace{-1mm}- \frac{1}{\eta^{\tau}}\ln \left(\sum_{k=1}^K\beta_k^{\tau}\cdot\exp\left\{-\eta^{\tau}P^{\tau}_k\right\}\hspace{-1mm}\right)\hspace{-1mm}\right)\hspace{-1mm}\right).
\end{align}
The quantity $\Delta^t$ in \eqref{eq_Delta_DCRC} represents the difference between two weighted average FPR objectives. The first, $\sum_{k=1}^K \beta^{\tau}_kP^{\tau}_k$, represents an upper bound on the objective \eqref{subeq_FPR}, while the second, $\ln \big(\sum_{k=1}^K\beta_k^{\tau}\cdot\exp\left\{-\eta^{\tau}P^{\tau}_k\right\}\big)/\eta^{\tau}$, is a strongly convex approximation \cite{vovk2001competitive, cesa2006prediction}. A larger value of $\Delta^t$ indicates a larger accumulated approximation error, which is addressed in the update \eqref{eq_wei_up_DCRC} by reducing the learning rate \eqref{eq_lr_up_DCRC}.

\subsection{Theoretical Guarantees}
\begin{algorithm}[t]
  \caption{D-CRC \cite{gasparin2024conformal}}\label{algo_D_CRC}
  \hspace*{\algorithmicindent}\parbox[t]{\dimexpr\linewidth-\algorithmicindent}{\textbf{Input:} Data stream $\left\{\left(X^t,Y^t\right)\right\}_{t=1}^T$, pre-trained machine learning models $\left\{g_k\left(\cdot\right)\right\}_{k=1}^K$, update step size $\rho$, target long-term FNR tolerance $\alpha$}
  \begin{algorithmic}[1]
    \STATE {\textbf{Initialization:} Initialize the common local threshold $\lambda^1=0$ at time $t=1$; set the global threshold $\theta^t=\theta\in(0,1]$ for all time $t$; set combining weights $\beta_k^1=1/K$ for $k=1,\ldots, K$ at time $t=1$; initialize cumulative error $\Delta^0=0$ at time $t=0$}
    \FOR{each time step $t$}
        \FOR{each sensor $k$}
            \STATE{Obtain a local observation $X^t_k$ of the state $X^t$}
            \STATE{Generate local probability vector $S^t_k$ via \eqref{eq_loc_prob_vec}}
            \STATE{Calculate local hard prediction $U^t_k$ via \eqref{eq_loc_pred_vec} and send it, uncompressed, to the central server}
        \ENDFOR
    \STATE{\textbf{At the central server:}}
    \STATE{\quad Calculate global probability vector $R^t$ via \eqref{eq_glo_prob_vec}}
    \STATE{\quad Calculate global prediction vector $V^t$ via \eqref{eq_glo_pred_vec}}
    \STATE{\quad Update the common local threshold $\lambda^t$ via rule \eqref{eq_loc_thre_up_DCRC}}
    \STATE{\quad Update the combining weights $\left\{\beta_k^t\right\}_{k=1}^K$ via rule \eqref{eq_wei_up_DCRC}}
    \ENDFOR
  \end{algorithmic}
\end{algorithm}

Overall, as summarized in Algorithm \ref{algo_D_CRC}, D-CRC applies the update rules for local thresholds and combining weights in \eqref{eq_loc_thre_up_DCRC} and \eqref{eq_wei_up_DCRC}. By following the analysis in \cite{gasparin2024conformal}, D-CRC can be proven to achieve the following guarantees for FPR objective in \eqref{subeq_FPR} and FNR constraint in \eqref{subeq_FNR}.

\begin{theorem}\label{theo_sim}
    For any $T\geq 1$, given an FNR target $\alpha\in[0,1]$, the time-averaged global FNR achieved by D-CRC is upper bounded by
    \begin{align}\label{eq_glo_FNR_ub_sim}
        \frac{1}{T}\sum_{t=1}^TN^t \leq \alpha+\frac{\lambda^1+\rho(1-\alpha)}{\rho T},
    \end{align}
    where $\lambda^1$ is the initial local threshold and $\rho$ is the learning rate in \eqref{eq_loc_thre_up_DCRC}. Furthermore, the time-averaged global FPR achieved by D-CRC is upper bounded by
    \begin{align}\label{eq_glo_FPR_ub_D_CRC}
        \frac{1}{T} \sum_{t=1}^TP^t\leq \frac{\mathcal{P}^*}{\theta} + \epsilon,
    \end{align}
    where $\mathcal{P}^*$ is the long-term local FPR of the \textit{best} sensor in hindsight, i.e., the sensor with the minimum long-term local FPR,
    \begin{align}\label{eq_opt_FPR}
        \mathcal{P}^* = \min\limits_{k\in\left\{1,\ldots,K\right\}} \frac{1}{T}\sum_{t=1}^T P^t_k.
    \end{align}
    The error $\epsilon$ in \eqref{eq_glo_FPR_ub_D_CRC} is given as
    \begin{align}\label{eq_FPR_err_D_CRC}
        \epsilon =& \frac{2}{T\theta}\left(\ln K\cdot\frac{\max\limits_{t\in\left\{1,\ldots, T\right\}} \hspace{-1mm}\Delta P^t}{\sum_{t=1}^T\Delta P^t}\right)^{\hspace{-1mm}\frac{1}{2}}\nonumber\\
        &\cdot \left(\sum_{t=1}^T P^t_{\text{\rm{max}}}-\hspace{-2mm}\min\limits_{k\in\left\{1,\ldots,K\right\}}\sum_{t=1}^TP_k^t\right)^{\hspace{-1mm}\frac{1}{2}}\nonumber\\
        &\cdot\left(\min\limits_{k\in\left\{1,\ldots,K\right\}}\sum_{t=1}^TP_k^t-\hspace{-1mm}\sum_{t=1}^TP^t_\text{\rm{min}}\right)^{\hspace{-1mm}\frac{1}{2}}
        \nonumber\\
        &+\frac{1}{T\theta}\left(\frac{16}{3}\ln K+2\right)\hspace{-1mm}\max_{t\in\left\{1,\ldots,T\right\}}\hspace{-1mm}\Delta P^t,
    \end{align}
    where $P^t_\text{\rm{max}}$ and $P^t_\text{\rm{min}}$ are the maximum and minimum local FPRs across $K$ sensors at time $t$, i.e.,
    \begin{align}\label{eq_FPR_max_min}
        P^t_\text{\rm{max}}=\max_{k\in\left\{1,\ldots,K\right\}}P^t_k~~\text{and}~~
        P^t_\text{\rm{min}}=\min\limits_{k\in\left\{1,\ldots,K\right\}}P^t_k,
    \end{align}
    respectively; and $\Delta P^t$ is the dynamic range of the local FPRs, i.e.,
    \begin{align}\label{eq_range_FPR}
        \Delta P^t = P^t_\text{\rm{max}} - P^t_\text{\rm{min}}.
    \end{align}
\end{theorem}
By \eqref{eq_glo_FNR_ub_sim}, D-CRC meets the FNR constraint \eqref{subeq_FNR}, while achieving a cumulative error $\epsilon$ with respect to the best-sensor FPR performance $\mathcal{P}^*$ of the order $\max\limits_{t\in\left\{1,\ldots, T\right\}}\Delta P^t\sqrt{T}$ in the worst-case scenario, exhibiting sublinear growth with $T$ \cite{de2014follow}. Consequently, Theorem \ref{theo_sim} states that D-CRC can meet the FNR constraint \eqref{subeq_FNR} while achieving approximately $1/\theta$ times the long-term FPR as compared to selecting the best sensor in hindsight after observing the true target $Y^t$.

\section{Communication-Constrained Distributed Conformal Risk Control}\label{sec_CD_CRC}
As discussed in the previous section, D-CRC tackles the problem \eqref{eq_obj} by neglecting the presence of the communication constraint \eqref{subeq_capacity}. When the communication constraint \eqref{subeq_capacity} is imposed, it is no longer possible to adapt the local thresholds $\lambda^t_k$ solely to meet the reliability constraint \eqref{subeq_FNR}, as the choice of the local thresholds $\lambda^t_k$ must also account for the bandwidth $B^t_k$ required to communicate the local decision $U^t_k$ to the server.

This section proposes communication-constrained distributed CRC (CD-CRC), a novel protocol that addresses problem \eqref{eq_obj} in the presence of communication constraint. By design, CD-CRC inherits the performance guarantees of D-CRC when the channel capacity is sufficiently large to make the capacity constraint \eqref{subeq_capacity} immaterial, while being able to address problem \eqref{eq_obj} even in the presence of stringent capacity constraints.

Unlike D-CRC, CD-CRC updates the local thresholds $\left\{\lambda^t_k\right\}_{k=1}^K$ in \eqref{eq_loc_pred_vec} by accounting for both constraints \eqref{subeq_FNR} and \eqref{subeq_capacity}. However, the limitations imposed by the communication constraint \eqref{subeq_capacity} prevent the updates of the local thresholds from being able to control the reliability requirement \eqref{subeq_FNR} by themselves. To obviate this problem, DC-CRC integrates the local threshold updates with the global threshold update.

\subsection{Local Threshold Update}\label{sec_loc_thre_up}
Optimization over the local thresholds $\left\{\lambda^t_k\right\}_{k=1}^K$ should ideally address both the reliability constraint \eqref{subeq_FNR}, as D-CRC, and the communication capacity constraint \eqref{subeq_capacity}. In particular, given the long-term channel capacity constraint $C$ in \eqref{subeq_capacity}, one should preferably allocate larger bandwidths to sensors communicating more informative decisions.

To this end, at each time $t$, DC-CRC first allocates \textit{target capacities} $\left\{C^t_k\right\}_{k=1}^K$ across the $K$ sensors, where $C^t_k\in[0,K]$ is the fractional capacity allocated to sensor $k$. The capacity $C^t_k$ serves as a target allocation for sensor $k$ at time $t$, and the allocations are chosen so as to meet the capacity constraint $C$ in \eqref{subeq_capacity} at each time $t$, i.e.,
\begin{align}\label{eq_cap_sum}
    \sum_{k=1}^K C^t_k = C.
\end{align}
While other choices are possible, one reasonable way to allocate capacities is to set
\begin{align}\label{eq_cap_alloc}
    C^t_k = \beta^t_k C,
\end{align}
where $\beta^t_k$ are the weights used for combining in \eqref{eq_glo_prob_vec}. The rationale for this choice is that the weight $\beta^t_k$ should be larger for sensors that have higher-quality observations. Note that other functions of the weights $\beta_k^t$ could also be used in lieu of \eqref{eq_cap_alloc}, as long as a higher capacity is assigned to sensor $k$ with a larger value $\beta_k^t$.

Given any capacity allocation \eqref{eq_cap_sum}, such as \eqref{eq_cap_alloc}, CD-CRC updates the local threshold at each sensor $k$ as
\begin{align}\label{eq_loc_thre_up}
    &\lambda^{t+1}_k = \lambda^t_k-\gamma\Delta\lambda^t_k~~\text{with}~~\nonumber\\
    &\Delta\lambda^t_k=
    \begin{cases}
	B^t_k-C^t_k & \text{if $B^t_k>C^t_k$},\\
        \max\left\{N^t-\alpha,B^t_k-C^t_k\right\} & \text{if $B^t_k\leq C^t_k$, $\lambda^t_k\geq -\gamma$},\\
        0 & \text{if $B^t_k\leq C^t_k$, $\lambda^t_k< -\gamma$},
    \end{cases}
\end{align}
where $\gamma>0$ denotes the step size. By \eqref{eq_loc_thre_up}, if the current communication cost $B^t_k$ exceeds the target $C^t_k$, the local threshold $\lambda^{t+1}$ is decreased to reduce the communication cost required at the next time slot so as to better meet the communication constraint \eqref{subeq_capacity}. Conversely, if the communication resources $C^t_k$ allocated to sensor $k$ at time $t$ are sufficient, i.e., $B^t_k\leq C^t_k$, the update \eqref{eq_loc_thre_up} considers also the FNR constraint \eqref{subeq_FNR} in a similar manner to the update \eqref{eq_loc_thre_up_DCRC} of D-CRC. The last condition in \eqref{eq_loc_thre_up} prevents the local threshold from becoming smaller than $-2\gamma$. This turns out to be useful in establishing the convergence of CD-CRC.

With regard to the adaptive update strategy \eqref{eq_loc_thre_up}, it is noted that switching between capacity and FNR constraints based on the communication overhead as in \eqref{eq_loc_thre_up} is effective both in targeting the capacity constraint \eqref{subeq_capacity} and in facilitating the attainment of the FNR constraint \eqref{subeq_FNR}. In particular, the second case in \eqref{eq_loc_thre_up} prevents the sensors from producing all-zero vector predictions $U^t_k$ when communication resources are not limiting, which would lead to a high local FNR.

\subsection{Global Threshold Update}\label{sec_glo_thre_up}
The control of the FNR enabled by the second case in the update \eqref{eq_loc_thre_up} of the local threshold is generally insufficient to meet the FNR constraint \eqref{subeq_FNR}. This is because unlike D-CRC, CD-CRC leverages the update \eqref{eq_loc_thre_up} also to ensure the communication constraint \eqref{subeq_capacity}. Therefore, in order to satisfy the long-term FNR constraint in \eqref{subeq_FNR}, CD-CRC applies an online CRC update to the global threshold $\theta^t$ based on the feedback on the FNR accrued by the previous decision $V^t$ as
\begin{align}\label{eq_update_glo}
    \tilde{\theta}^{t+1}=\tilde{\theta}^t-\mu\left(N^t-\alpha\right),
\end{align}
and
\begin{align}\label{eq_glo_thre_offset}
    \theta^{t+1} = \tilde{\theta}^{t+1} - \delta,
\end{align}
where $\mu>0$ represents the step size and $\delta>0$ is a hyperparameter. In words, by the first term on the right-hand side of \eqref{eq_update_glo}, the global threshold $\theta^{t+1}$ is decreased if the current FNR fails to meet the constraint $\alpha$, while it is increased otherwise. Given the global decision \eqref{eq_glo_pred_vec}, this implies that in the former case, the number of labels deemed to be relevant will tend to increase, reducing the FNR at the next iteration $t+1$; while, in the latter case, the FNR will tend to grow at the next time $t+1$. The need for the correction $\delta>0$ in \eqref{eq_glo_thre_offset} will be made clear by the analysis below. In this regard, it is noted here that setting the correction $\delta$ such that
\begin{align}\label{eq_lb_hyp_glo_pre}
    \delta > \mu(1-\alpha)
\end{align}
ensures that the corrected global threshold $\tilde{\theta}^t$ in \eqref{eq_glo_thre_offset} is strictly positive for all time $t$. For a detailed proof of this claim, please refer to Appendix \ref{apdx_range_glo_thre}.

\subsection{Combining Weights Update}\label{sec_wei_up}
Sec. \ref{sec_FNR_analysis} and Sec. \ref{Sec_commu_analysis} will demonstrate that the updates \eqref{eq_update_glo} and \eqref{eq_loc_thre_up} guarantee the constraints \eqref{subeq_FNR} and \eqref{subeq_capacity}, respectively. Therefore, the combining weights $\left\{\beta^t_k\right\}_{k=1}^K$ can be updated solely with the aim of minimizing the long-term FPR in \eqref{subeq_FPR} without needing to address the constraints in problem \eqref{eq_obj}.

To this end, as in the D-CRC combining rule \eqref{eq_wei_up_DCRC}, we adopt the \textit{online exponentiated gradient rule} \cite{orabona2019modern, gasparin2024conformal, de2014follow} updating the combining weights as
\begin{align}\label{eq_wei_up}
    \beta^{t+1}_k=\frac{\exp\left\{-\eta^{t+1}\sum_{\tau=1}^tP^{\tau}_k/\tilde{\theta}^{\tau}\right\}}{\sum_{k=1}^K\exp\left\{-\eta^{t+1}\sum_{\tau=1}^tP^{\tau}_k/\tilde{\theta}^{\tau}\right\}},
\end{align}
with adaptive learning rate $\eta^t$. As compared to the D-CRC rule \eqref{eq_wei_up_DCRC}, the update \eqref{eq_wei_up} scales each term $P^{\tau}_k$ by the inverse of the corrected global threshold, $1/\tilde{\theta}^\tau$. This way, the threshold $\tilde{\theta}^\tau$ serves as a temperature parameter: As $\tilde{\theta}^{\tau}\rightarrow 0$, the sensor with the smallest FPR tends to be assigned weight equal to $1$, while other sensors have near-zero weights, while, as $\tilde{\theta}^{\tau}\rightarrow \infty$, sensors tend to be assigned uniform weights.

\begin{algorithm}[t]
  \caption{CD-CRC}\label{algo_CD_CRC}
    \hspace*{\algorithmicindent}\parbox[t]{\dimexpr\linewidth-\algorithmicindent}{\textbf{Input:} Data stream $\left\{\left(X^t,Y^t\right)\right\}_{t=1}^T$, pre-trained machine learning models $\left\{g_k\left(\cdot\right)\right\}_{k=1}^K$, local update step size $\gamma$, global update step size $\mu$, long-term FNR tolerance $\alpha$, long-term capacity constraint $C$}
  \begin{algorithmic}[1]
    \STATE {\textbf{Initialization:} Initialize the local thresholds $\lambda_k^1=0$ for $k=1,\ldots,K$ and global threshold $\theta^1=0$ at time $t=1$; set combining weights $\beta_k^1=1/K$ for $k=1,\ldots, K$ at time $t=1$; initialize cumulative error $\Gamma^0=0$ at time $t=0$}
    \FOR{each time step $t$}
        \FOR{each sensor $k$}
            \STATE{Obtain a local observation $X^t_k$ of the state $X^t$}
            \STATE{Generate local probability vector $S^t_k$ via \eqref{eq_loc_prob_vec}}
            \STATE{Calculate local hard prediction $U^t_k$ via \eqref{eq_loc_pred_vec}}
            \STATE{Update local threshold $\lambda_k^t$ via rule \eqref{eq_loc_thre_up} and transmit it to the central server}
        \ENDFOR
    \STATE{\textbf{At the central server:}}
    \STATE{\quad Calculate global probability vector $R^t$ via \eqref{eq_glo_prob_vec}}
    \STATE{\quad Calculate global prediction vector $V^t$ via \eqref{eq_glo_pred_vec}}
    \STATE{\quad Update global threshold $\theta^t$ via rule \eqref{eq_update_glo}}
    \STATE{\quad Update the combining weights $\left\{\beta_k^t\right\}_{k=1}^K$ via rule \eqref{eq_wei_up}}
    \ENDFOR
  \end{algorithmic}
\end{algorithm}

Finally, in a similar manner to \eqref{eq_lr_up_DCRC}, the learning rate in \eqref{eq_wei_up} is updated as
\begin{align}\label{eq_lr_up}
    \eta^{t+1}=\frac{\ln K}{\Gamma^t},
\end{align}
where
\begin{align}\label{eq_Delta}
    \Gamma^t = \sum_{\tau=1}^t\left(\sum_{k=1}^K \frac{\beta^{\tau}_k}{\tilde{\theta}^{\tau}}P^{\tau}_k + \frac{1}{\eta^{\tau}}\ln \left(\sum_{k=1}^K\beta_k^{\tau}\cdot\exp\left\{\frac{-\eta^{\tau}}{\tilde{\theta}^{\tau}}P^{\tau}_k\right\}\right)\right).
\end{align}

\subsection{Summary of CD-CRC}
The overall workflow of the proposed CD-CRC protocol is detailed in Algorithm \ref{algo_CD_CRC}. At each time $t$, given an input $X^t$, each sensor $k$ obtains a noisy observation $X^t_k$ and inputs it into the local pre-trained model $g_k(\cdot)$ to generate the local probability vector $S^t_k$ as described in \eqref{eq_loc_prob_vec}. The local prediction vector $U^t_k$ is then generated by using the thresholding scheme in \eqref{eq_loc_pred_vec}. Each sensor $k$ updates its local threshold $\lambda^t_k$ via the rule \eqref{eq_loc_thre_up} to satisfy the long-term capacity constraint \eqref{subeq_capacity}. Subsequently, each sensor $k$ transmits the local prediction vector $U^t_k$ to the central server for final decision.

At the central server, the $K$ local predictions are aggregated by using a weighted combination scheme to obtain the global probability vector $R^t$ via \eqref{eq_glo_prob_vec}. The global prediction vector $V^t$ is then derived through the thresholding scheme in \eqref{eq_glo_pred_vec}. With feedback on the true decision $Y^t$, the central server updates the global threshold $\theta^t$ by using \eqref{eq_update_glo} and \eqref{eq_glo_thre_offset}, and it updates the combining weights $\left\{\beta^t_k\right\}_{k=1}^K$ by using the online exponentiated gradient rule \eqref{eq_wei_up}, aiming to minimize the long-term FPR in \eqref{subeq_FPR} while satisfying the long-term FNR tolerance in \eqref{subeq_FNR}.

\section{Theoretical Analysis}\label{sec_theo}
In this section, we analyze the performance of the proposed CD-CRC protocols in relation to problem \eqref{eq_obj}. First, we demonstrate that, under mild assumptions, CD-CRC satisfies the FNR and capacity constraints specified in \eqref{subeq_FNR} and \eqref{subeq_capacity}, respectively. Then, we evaluate the long-term FPR in \eqref{subeq_FPR} that is achieved by CD-CRC in comparison to the long-term FPR achieved by the best sensor in hindsight.

\subsection{Long-term FNR Constraint}\label{sec_FNR_analysis}
We first analyze the achievability of the long-term global FNR constraint in \eqref{subeq_FNR} via CD-CRC.
\begin{theorem}\label{Theo_FNR}
    For any $T\geq 1$, the average global FNR achieved by CD-CRC is upper bounded as
    \begin{align}\label{eq_glo_FNR_ub}
        \frac{1}{T}\sum_{t=1}^T N^t < \alpha+\frac{\tilde{\theta}^1}{\mu T},
    \end{align}
    where $\tilde{\theta}^1$ is the initialization for the corrected global threshold \eqref{eq_update_glo} and $\mu>0$ is the learning rate in \eqref{eq_update_glo}.
\end{theorem}
\textit{Proof:} 
The proof, which follows closely \cite{feldman2022achieving}, is detailed in Appendix \ref{apdx_FNR_gua}.

According to \eqref{eq_glo_FNR_ub}, as $T\rightarrow \infty$, the long-term FNR constraint in \eqref{subeq_FNR} is satisfied. As anticipated in Sec. \ref{subsec_online}, the result \eqref{eq_glo_FNR_ub} can be extended to a setting with imperfect FNR feedback. To elaborate, suppose that, at each time $t$, the central server is only given a conservative estimate $\hat{N}^t\geq N^t$ of the FNR in \eqref{eq_FNR}. Replacing $\hat{N}^t$ for $N^t$ in the update \eqref{eq_update_glo}, CD-CRC also guarantees the condition \eqref{eq_glo_FNR_ub}, as it follows directly from the proof in Appendix \ref{apdx_FNR_gua}.

\begin{figure*}[t]
    \centering
    {
    \includegraphics[width = 0.33\textwidth]{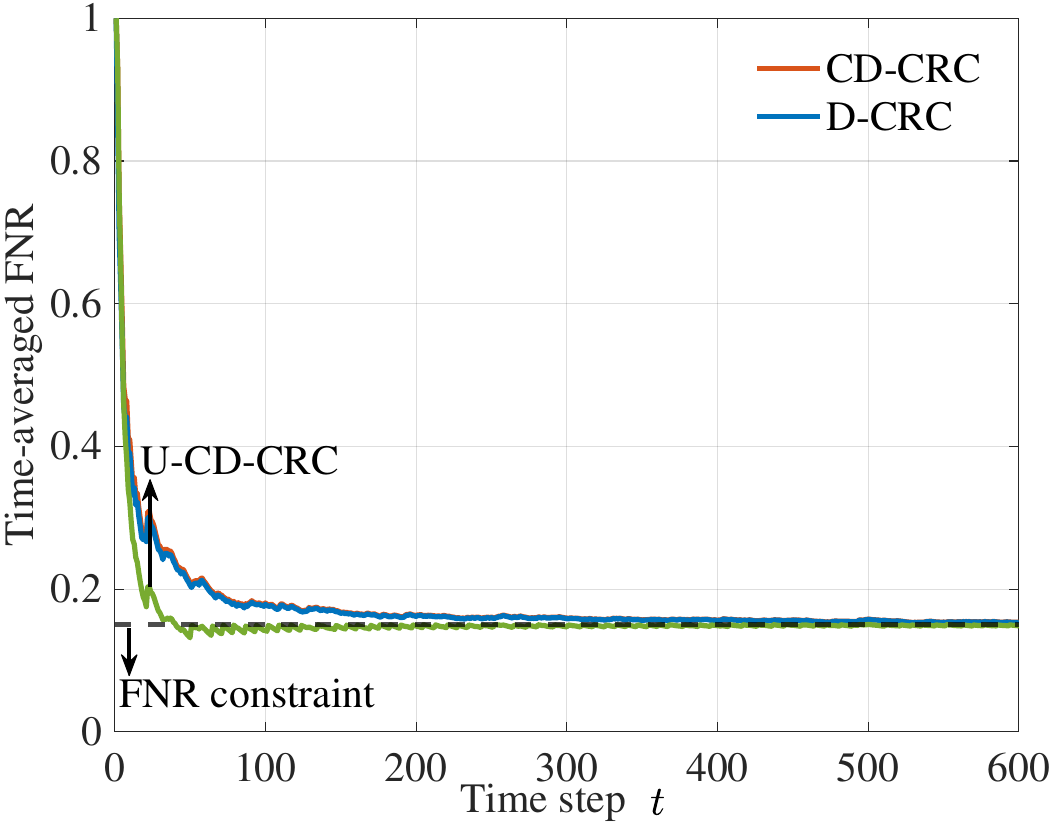}
    \includegraphics[width = 0.322\textwidth]{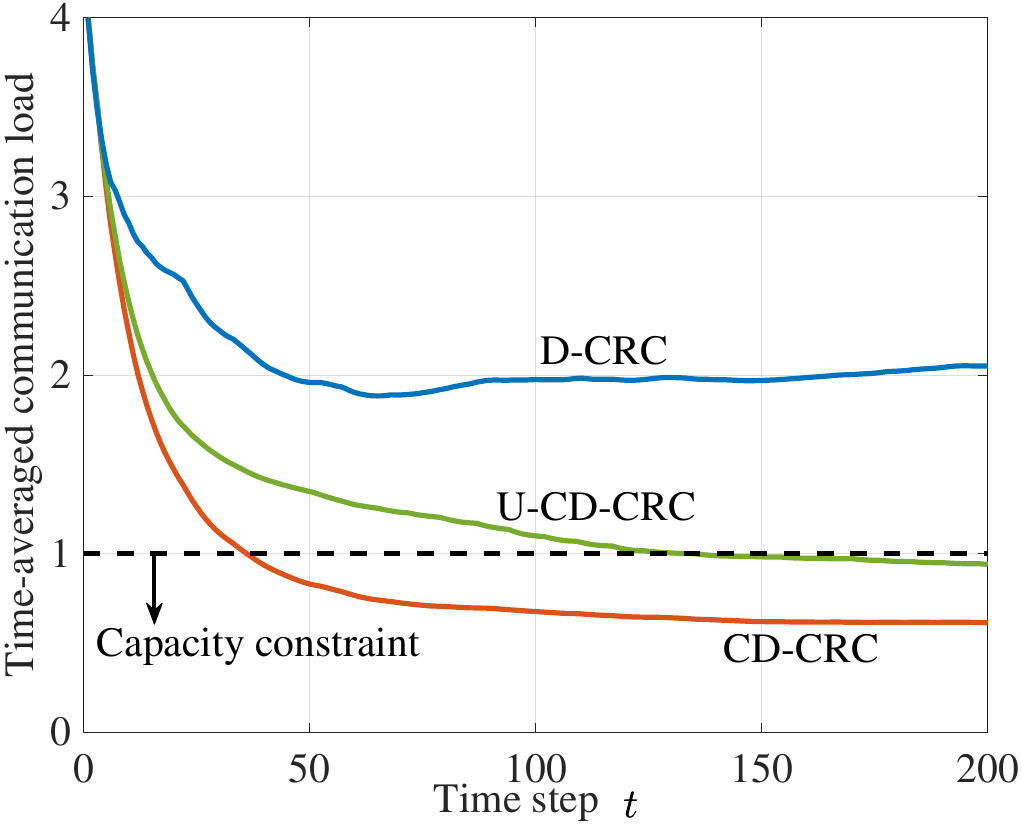}
    \includegraphics[width = 0.33\textwidth]{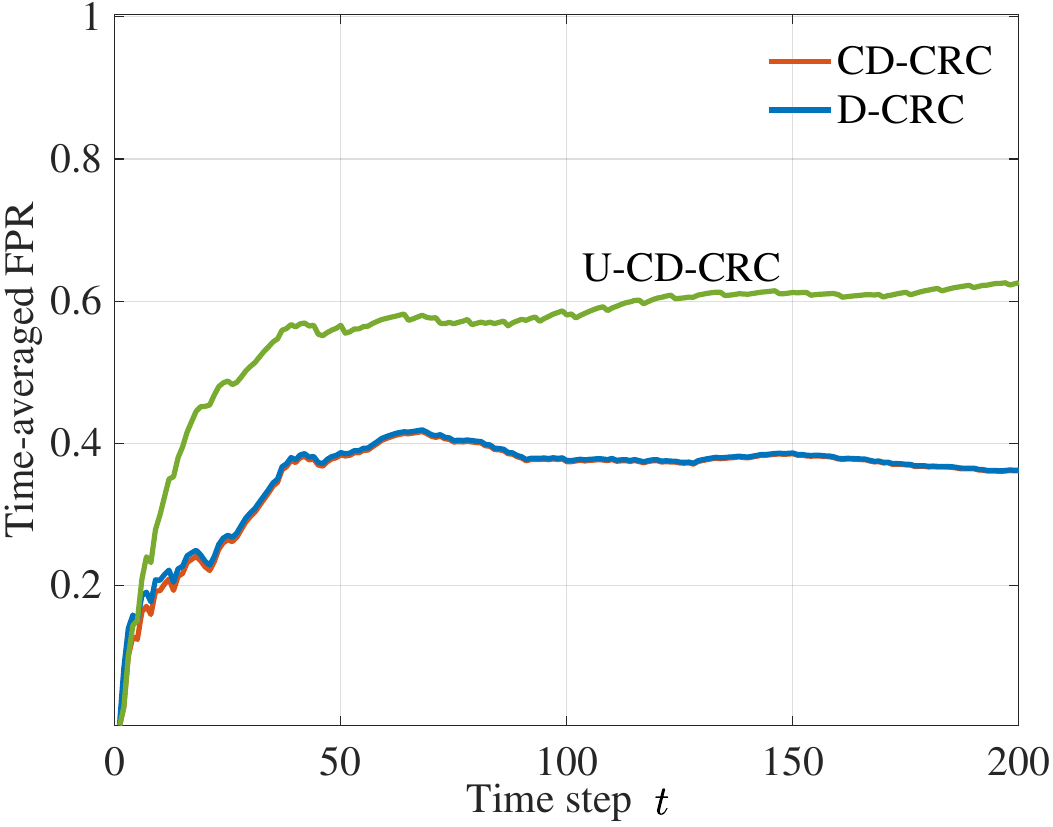}
    }
    \caption{Time-averaged FNR (left), time-averaged communication load (center), and time-averaged FPR (right) for CD-CRC, U-CD-CRC and D-CRC \cite{gasparin2024conformal} over time $t$ with FNR constraint $\alpha=0.15$ and capacity constraint $C=1$.}
    \label{performance_time}
\end{figure*}

\subsection{Long-term Communication Constraint}\label{Sec_commu_analysis}
We now address the capacity constraint \eqref{subeq_capacity}.
\begin{theorem}\label{Theo_cap}
    The long-term communication load required by CD-CRC is upper bounded as
    \begin{align}\label{eq_ub_capacity}
        \frac{1}{T}\sum_{t=1}^T\sum_{k=1}^K B^t_k
        \leq& C + \frac{\sum_{k=1}^K \left(\lambda^1_k+2\gamma\right)}{\gamma T},
    \end{align}
    where $\lambda^t_k$ is the initialization of the local threshold at sensor $k$ and $\gamma>0$ is the learning rate in \eqref{eq_loc_thre_up}.
\end{theorem}
\textit{Proof:} The proof is detailed in Appendix \ref{apdx_loc_LZ}.

By \eqref{eq_ub_capacity}, for any capacity allocation \eqref{eq_cap_alloc}, as $T\rightarrow\infty$, the capacity constraint in \eqref{subeq_capacity} is satisfied by CD-CRC.

\subsection{Long-term FPR Objective}\label{subsec_bound_FPR}
Finally, we establish an upper bound on long-term global FPR obtained by CD-CRC. The following lemma serves as a stepping stone for achieving the main result, as it elucidates the relation between local and global FPRs under the weighted combining rule \eqref{eq_wei_up}.
\begin{lemma}\label{lemma_up_FPR}
    At any time $t$, given any set of non-negative combining weights $\left\{\beta^t_k\right\}_{k=1}^K$, if the correction $\delta$ in \eqref{eq_glo_thre_offset} satisfies \eqref{eq_update_glo}, the objective $P^t$ in problem \eqref{eq_obj} achieved by CD-CRC is upper bounded as
        \begin{align}\label{eq_CDCRC_FPR_ub}
            P^t \leq \frac{1}{\tilde{\theta}^t}\sum_{k=1}^K\beta_k^t P^t_k + \frac{\delta}{\tilde{\theta}^t},
        \end{align}
    where $P^t_k$ represents the local FPR at the $k$-th sensor.
\end{lemma}
\textit{Proof:} 
The proof is detailed in Appendix \ref{apdx_FPR_ub}.

The result \eqref{eq_CDCRC_FPR_ub} illustrates the importance of introducing the correction in \eqref{eq_glo_thre_offset}, which ensures the non-negativity of the upper bound in \eqref{eq_CDCRC_FPR_ub}. Building on this lemma, we now present the main result for the average FPR attained by CD-CRC.
\begin{theorem}\label{Theo_FPR}
    The average global FPR over time $T$ is upper bounded by
    \begin{align}\label{eq_upbound2}
        \frac{1}{T} \sum_{t=1}^T P^t\leq \frac{\mathcal{P}^*}{\delta-\mu(1-\alpha)} + \sigma + \frac{\delta}{T}\sum_{t=1}^T\frac{1}{\tilde{\theta}^t},
    \end{align}
    where $\mathcal{P}^*$ denotes the minimum long-term local FPR incurred by the best sensor in hindsight as defined in \eqref{eq_opt_FPR}, and the error $\sigma$ is given by
    \begin{align}\label{eq_FPR_err_CD_CRC}
        \sigma =& \frac{2}{T}\left(\frac{\left(1+\delta+\mu\alpha\right)\ln K}{\delta-\mu(1-\alpha)}\cdot\frac{\max\limits_{t\in\left\{1,\ldots, T\right\}}\Delta P^t}{\sum_{t=1}^T\Delta P^t}\right)^{\frac{1}{2}}\nonumber\\
        &\cdot\left(\frac{\sum_{t=1}^TP^t_{\text{\rm{max}}}}{\delta-\mu(1-\alpha)}-\frac{\min\limits_{k\in\left\{1,\ldots,K\right\}}\sum_{t=1}^TP_k^t}{1+\delta+\mu\alpha}\right)^\frac{1}{2}\nonumber\\
        &\cdot\left(\frac{\min\limits_{k\in\left\{1,\ldots,K\right\}}\sum_{t=1}^TP_k^t}{\delta-\mu(1-\alpha)}-\frac{\sum_{t=1}^TP^t_\text{\rm{min}}}{1+\delta+\mu\alpha}\right)^{\frac{1}{2}}\nonumber\\
        &+\frac{1}{T(\delta-\mu(1-\alpha))} \left(\frac{16}{3}\ln K+2\right) \hspace{-1mm}\max\limits_{t\in\left\{1, \ldots, T\right\}}\hspace{-0.2cm}\Delta P^t,
    \end{align}
    where $P^t_\text{\rm{max}}$ and $P^t_\text{\rm{min}}$ are the maximum and minimum local FPRs across $K$ sensors at time $t$ as defined in \eqref{eq_FPR_max_min}; and $\Delta P^t$ is the dynamic range of the local FPRs as defined in \eqref{eq_range_FPR}.
\end{theorem}
\textit{Proof:} 
The proof is detailed in Appendix \ref{apdx_FPR_ub2}.

Since, in a similar manner to \eqref{eq_FPR_err_D_CRC}, the error $\sigma$ in \eqref{eq_FPR_err_CD_CRC} diminishes with increasing time $T$  \cite{gasparin2024conformal,de2014follow}, the result \eqref{eq_upbound2} shows that CD-CRC can approach an FPR equal to $1/(\delta-\mu(1-\alpha))$ times the smallest local long-term FPR $\mathcal{P}^*$ with an error proportional to $\delta$. Thus, decreasing the offset $\delta$ reduces the asymptotic error term in \eqref{eq_upbound2}, although this case at the expense of the multiplicative factor $1/\left(\delta-\mu\left(1-\alpha\right)\right)$. In practice, based on experimental evidence, a good realization is observed to be $\delta=\mu(1-\alpha)+\Delta\delta$, where $\Delta\delta$ is a small number that balances the multiplicative and addictive errors in \eqref{eq_upbound2}.

\begin{figure*}[t]
    \centering
    {
    \includegraphics[width = 0.3252\textwidth]{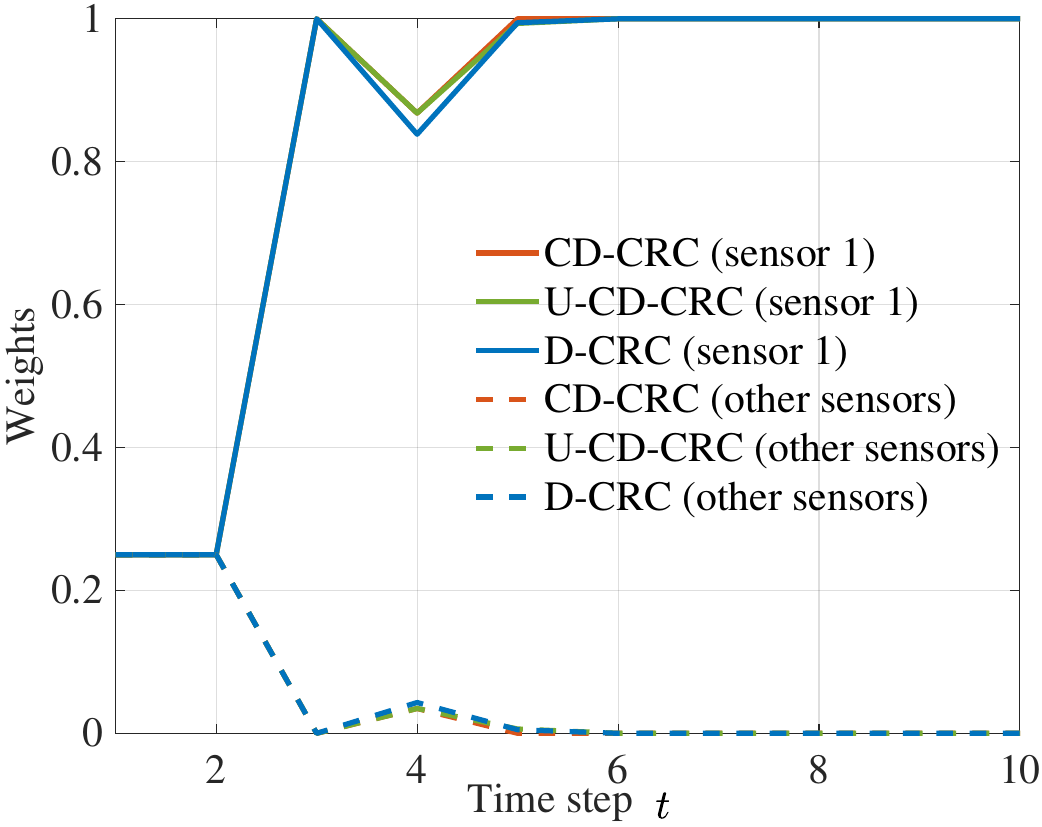}
    \includegraphics[width = 0.335\textwidth]{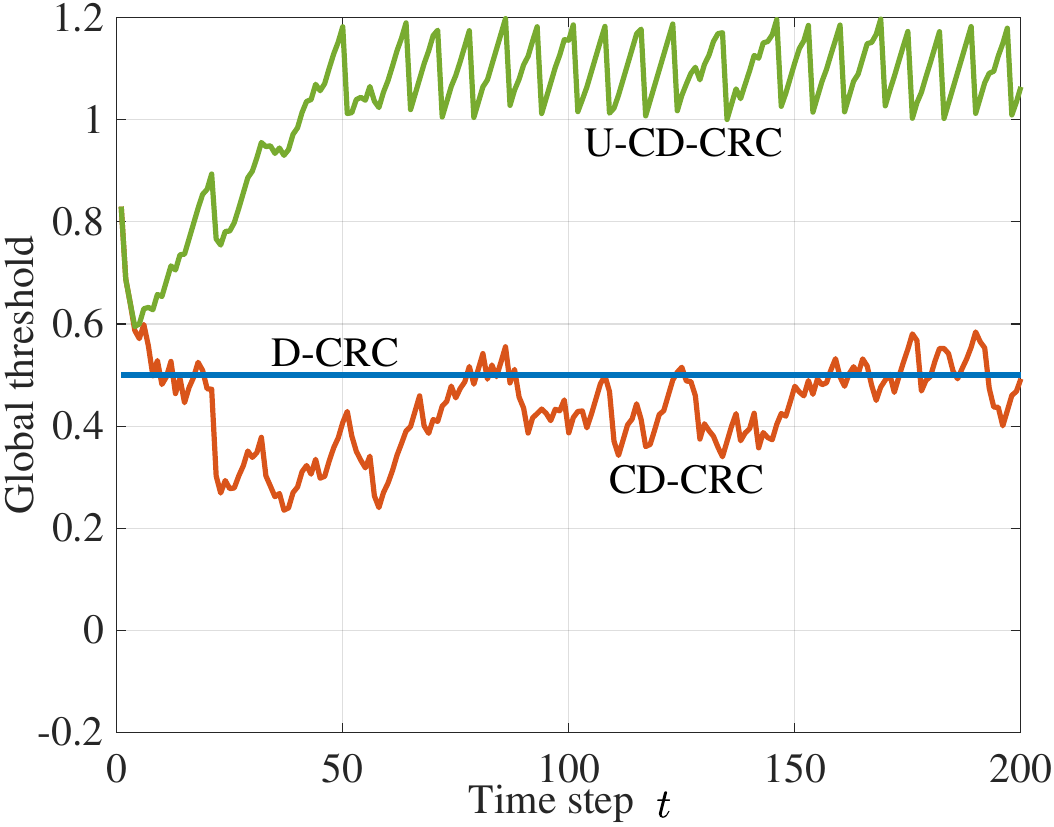}
    \includegraphics[width = 0.335\textwidth]{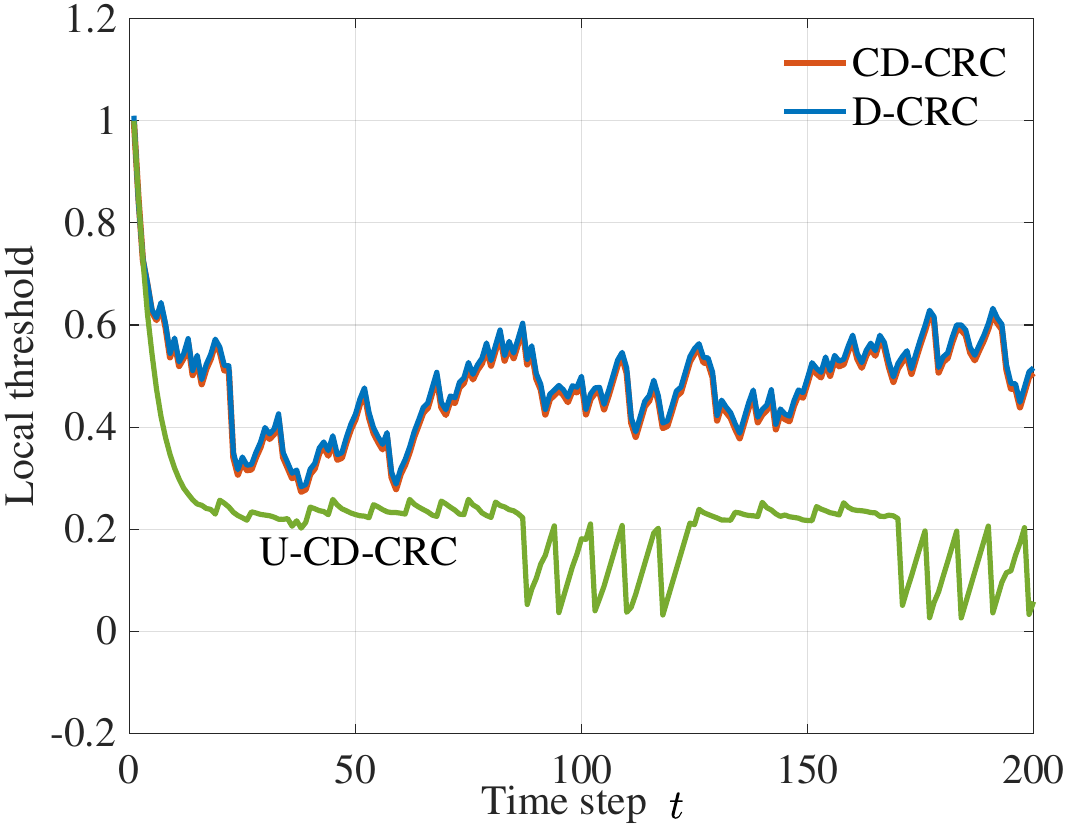}
    \includegraphics[width = 0.335\textwidth]{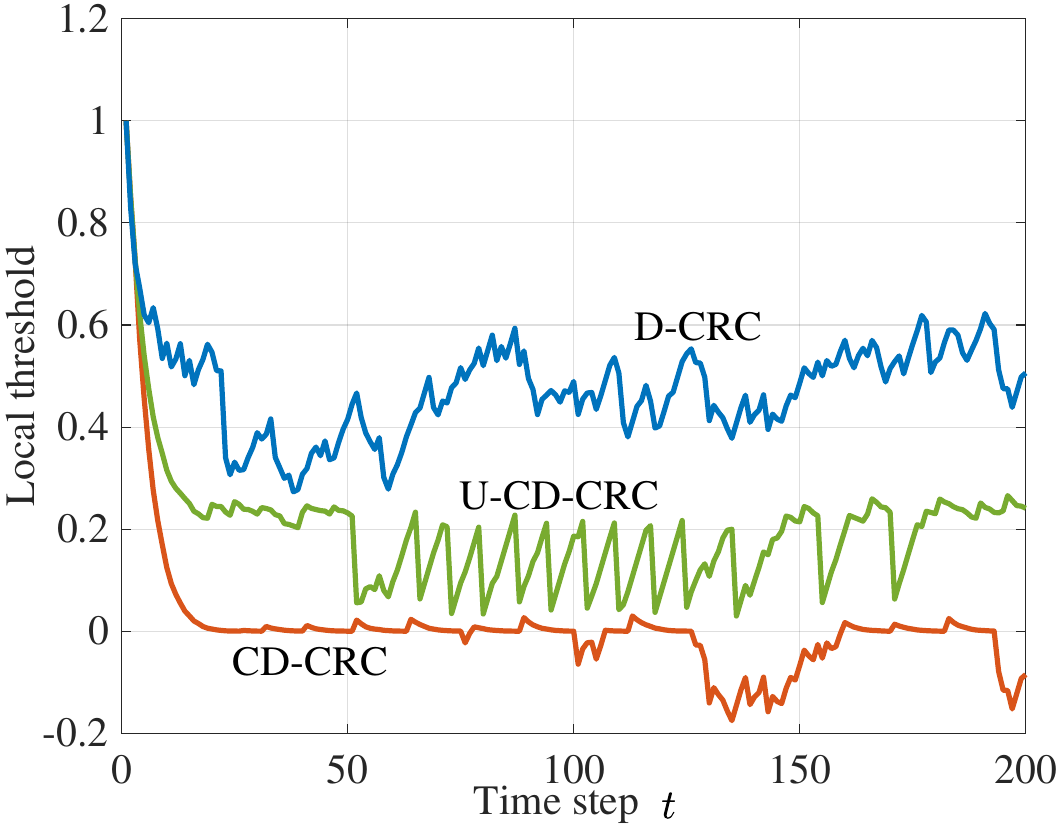}
    }
    \caption{Evolution of the weights (top-left corner), global thresholds (top-right corner), local thresholds for the optimal sensor (bottom-left corner), local thresholds for other non-optimal sensors (bottom-right corner) for CD-CRC, U-CD-CRC and D-CRC \cite{gasparin2024conformal} over time $t$ with FNR constraint $\alpha=0.15$ and capacity constraint $C=1$.}
    \label{performance_time2}
\end{figure*}

\section{Experimental Settings and Results}\label{sec_experiments}
In this section, we evaluate the performance of the proposed CD-CRC as compared to the benchmark D-CRC reviewed in Sec. \ref{sec_D_CRC}.

\subsection{Setting}\label{Sec_set}
We utilize the PASCAL VOC 2012 data set \cite{everingham2010pascal}, a standard benchmark for segmentation tasks, comprising $2913$ images with masks annotated with $20$ object classes and one background class. We pre-process the masks by grouping the $20$ object classes into one class labeled as ``object'', creating a binary segmentation task to identify objects (relevant) and background (irrelevant). We use $2000$ data pairs for training the predictive model and the remaining $913$ points for online testing. All images are resized to a uniform resolution of $100\times 100$ pixels.

We employ pre-trained fully convolutional networks \cite{long2015fully} with a ResNet-50 backbone \cite{he2016deep}, downloaded from the Torchvision repository. The last layer is modified to a convolutional layer followed by a softmax layer to produce the object probability vector $S^t_k$. The model is fine-tuned on the $2000$ training data points using cross-entropy as the loss function. We employ the SGD optimizer with a learning rate of $0.005$, momentum of $0.9$, and weight decay of $0.0001$ over $50$ epochs.

During online optimization, the step sizes $\rho$ in \eqref{eq_loc_thre_up_DCRC}, $\gamma$ in \eqref{eq_loc_thre_up}, and $\mu$ in \eqref{eq_update_glo} for local and global threshold updates are all set to $0.2$. The hyperparameter $\delta$ in the predictor \eqref{eq_glo_pred_vec} is set as $\delta = \mu(1-\alpha) + 0.01$ for CD-CRC to satisfy the requirement \eqref{eq_lb_hyp_glo_pre}.

For transmitting the local binary prediction vectors $U^t_k$ of length $L = 10000$, the data is divided into blocks of $10$ entries. Each block is encoded using a lossless compression scheme \cite{kontoyiannis2013optimal} satisfying the desired conditions explained in Sec. \ref{subsec_com_cons}. In this scheme, all possible $2^{10}$ binary messages of length $10$ are listed and ordered by the number of zeros, from fewest to most. Codewords are then assigned sequentially, starting with the shortest codewords, including empty codewords and all possible $1$-bit and $2$-bit combinations, continuing until every one of the $2^{10}$ messages has a unique codeword. Note that this approach meets the requirements \eqref{encode_require1} and \eqref{encode_require2}, as well as the property of requiring transmitted signal lengths $B^t_k L$ that increase with the number of zeros in the local decisions $U^t_k$.

To simulate real-world sensor limitations, we model scenarios where each sensor observes only a portion of the original object $X^t$, with unobserved areas filled with black pixels. We consider $K=4$ sensors, each generating observations through the following transformations:
\begin{itemize}
    \item \textbf{Base} ($X^t_1$): Sensor $1$ observes a random $80\times 80$ region of the original image $X^t$.
    \item \textbf{Brightness adjustment} ($X_2^t$): Sensor $2$ observes a random $60\times 60$ region of $X^t$ with a $20\%$ reduction in brightness.
    \item \textbf{Noise addition} ($X^t_3$): Sensor $3$ observes a random $60\times 60$ region of $X^t$ with added zero-mean Gaussian noise with a variance of $1000$.
    \item \textbf{Blur effect} ($X^t_4$): Sensor $4$ observes a random $60\times 60$ region of $X^t$ with a Gaussian blur of kernel size $7$.
\end{itemize}
For additional experiments conducted under different settings, please refer to Appendix \ref{apdx_experiment}.

We implement two versions of CD-CRC, one using the capacity allocation strategy \eqref{eq_cap_alloc}, which we label as CD-CRC; and one that distributes the available communication capacity evenly among the sensors by setting
\begin{align}\label{eq_even_alloc}
    C^t_k = \frac{C}{K}.
\end{align}
The latter, referred to as U-CD-CRC, serves as a natural benchmark to validate the effectiveness of the proposed capacity allocation strategy \eqref{eq_cap_alloc} for CD-CRC.

\begin{figure*}[t]
    \centering
    {\includegraphics[width = 0.32\textwidth]{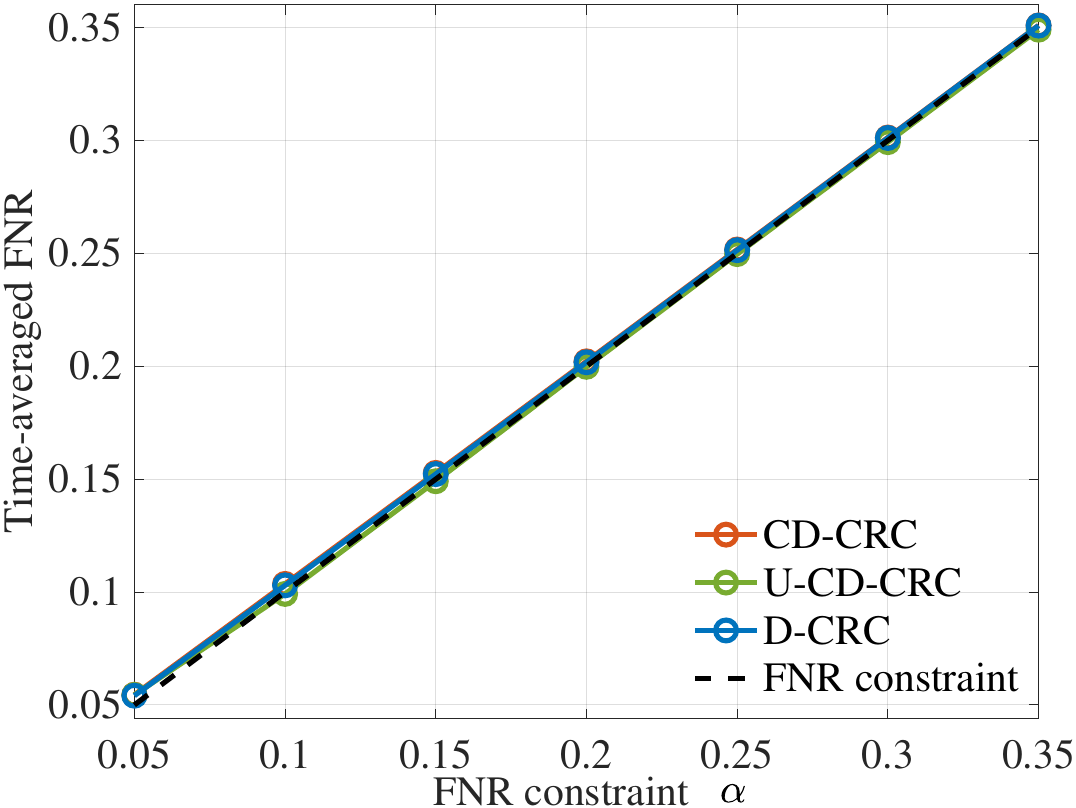}
    \includegraphics[width = 0.316\textwidth]{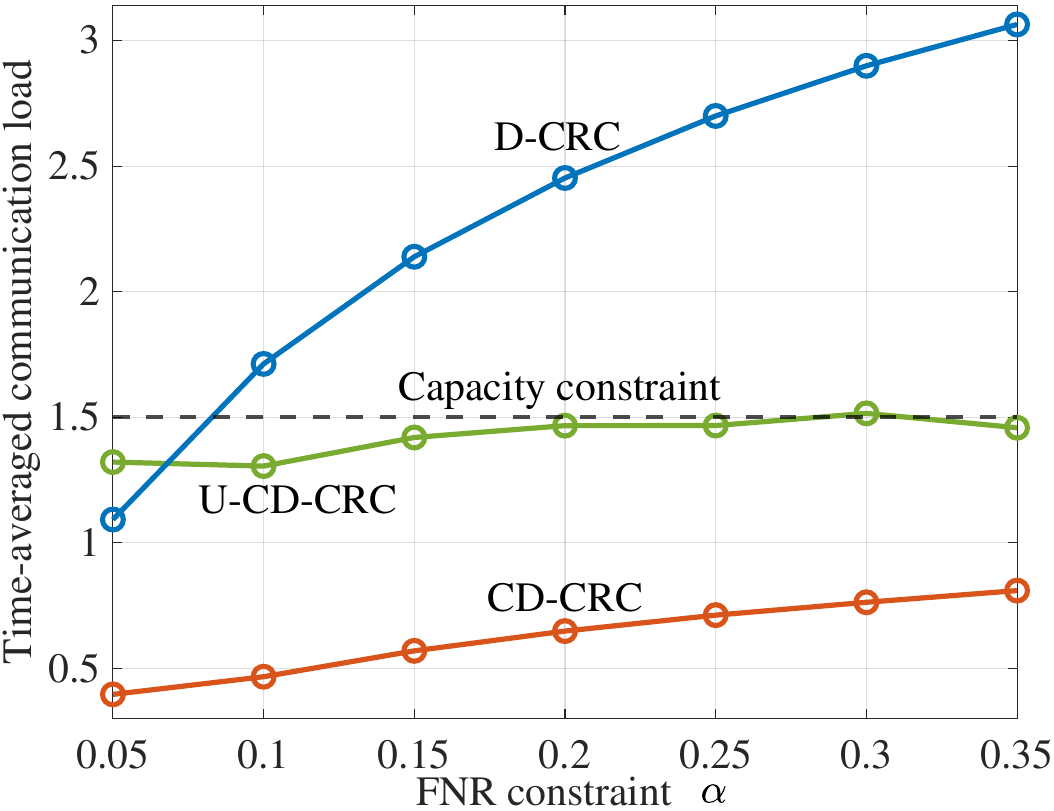}}
    \includegraphics[width = 0.32\textwidth]{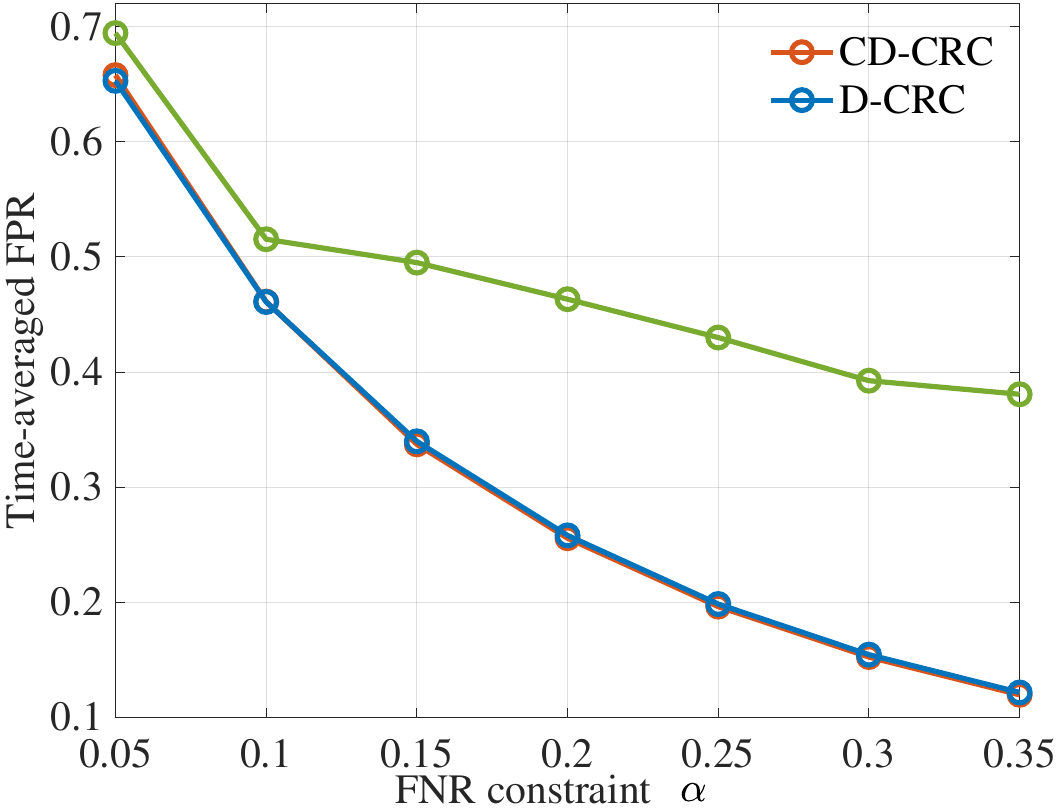}
    \caption{Time-averaged FNR (left), time-averaged communication load (middle), and time-averaged FPR (right) for CD-CRC, U-CD-CRC, and D-CRC \cite{gasparin2024conformal} versus the FNR constraint $\alpha$ with capacity constraint $C = 1.5$.}
    \label{longPer_alpha}
\end{figure*}

\begin{figure*}[t]
    \centering
    {
	\includegraphics[width = 0.32\textwidth]{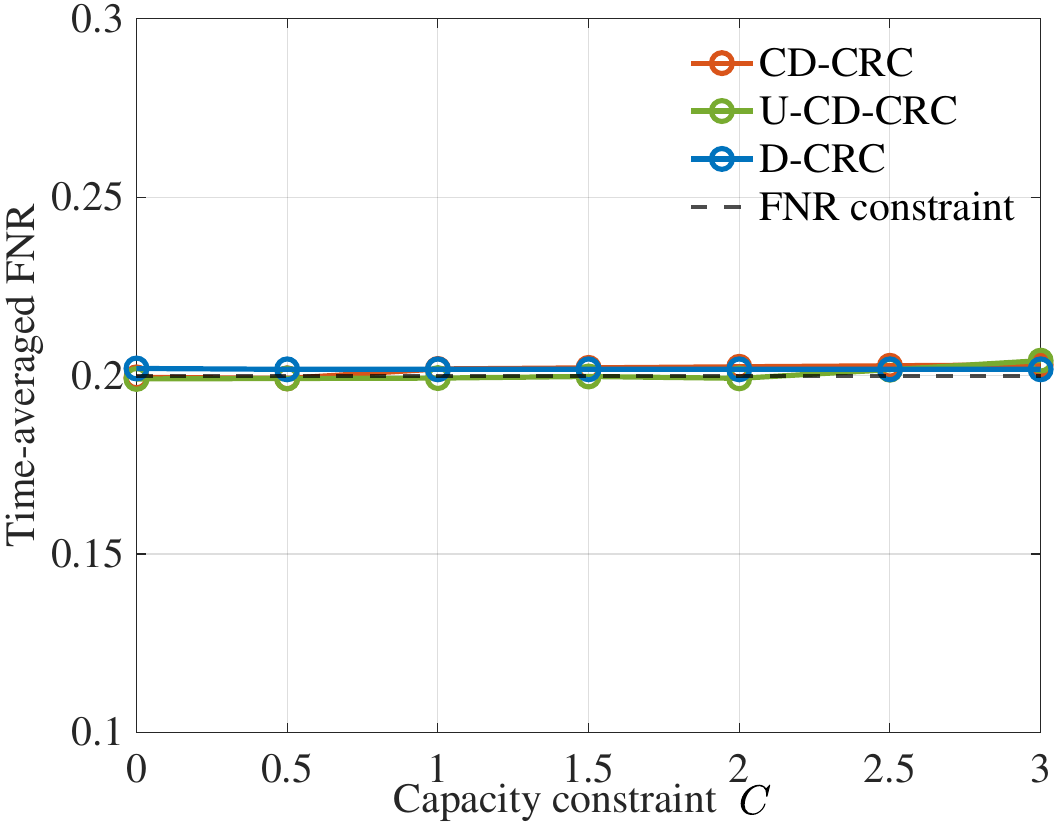}
    \includegraphics[width = 0.315\textwidth]{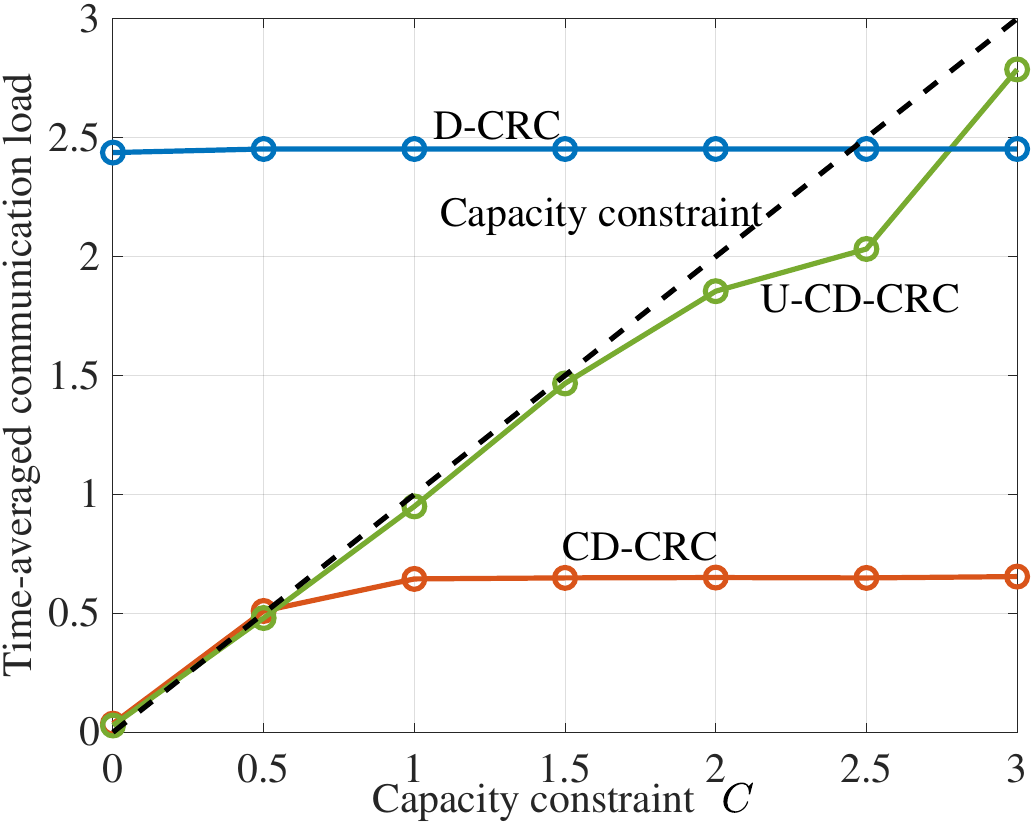}}
    \includegraphics[width = 0.32\textwidth]{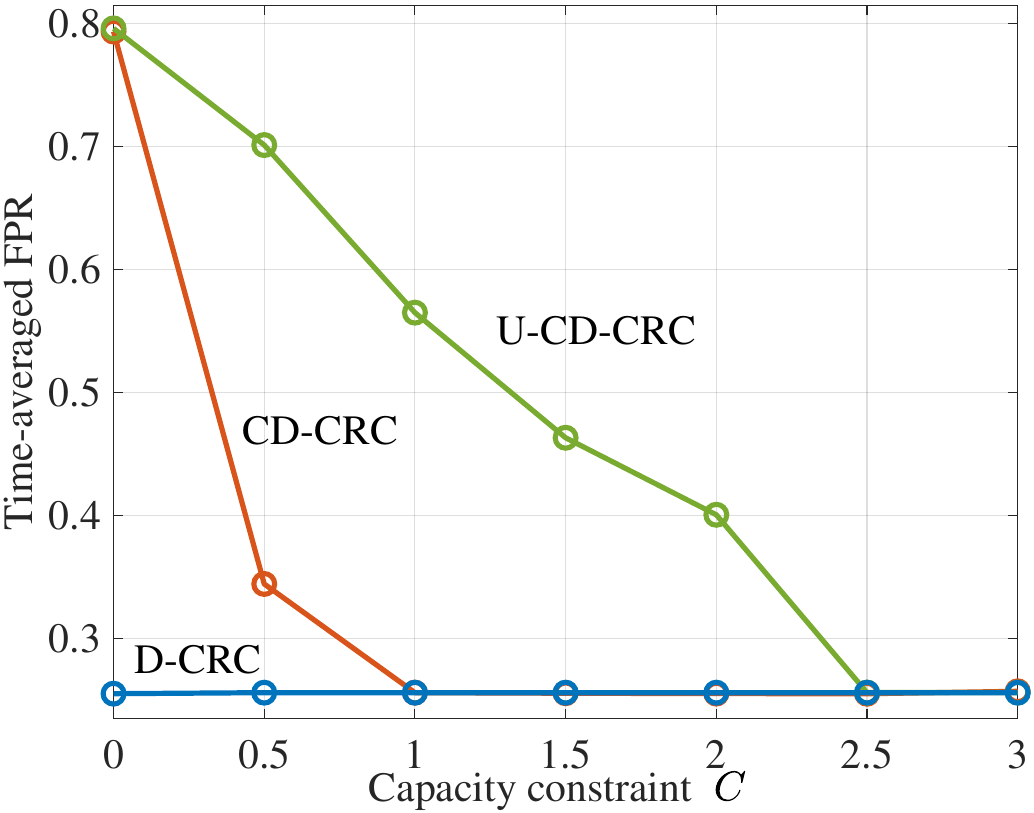}
    \caption{Time-averaged FNR (left), time-averaged communication load (middle), and time-averaged FPR (right) for CD-CRC, U-CD-CRC, and D-CRC \cite{gasparin2024conformal} versus the capacity constraint $C$ with FNR constraint $\alpha = 0.2$.}
    \label{longPer_C}
\end{figure*}

\subsection{Performance Evaluation}\label{subsec_performance}
To start, we evaluate the performance of D-CRC \cite{gasparin2024conformal}, as well as of the two mentioned versions of the proposed CD-CRC, as a function of the time step $t$. The performance is averaged over $200$ realizations of sensor scenarios as described in Sec. \ref{Sec_set}. We focus on the time-averaged FNR, $\sum_{\tau=1}^tN^{\tau}/t$ with $N^{\tau}$ in \eqref{eq_FNR}, the time-averaged communication load $\sum_{\tau=1}^t\sum_{k=1}^KB^{\tau}_k/t$, and the time-averaged FPR $\sum_{\tau=1}^tP^{\tau}/t$ with $P^{\tau}$ in \eqref{eq_FPR}, which are shown in Fig. \ref{performance_time}. We also plot the evolution of the weights $\left\{\beta^t_k\right\}_{k=1}^K$, the local thresholds $\left\{\lambda^t_k\right\}_{k=1}^K$, and the global threshold $\theta^t$ in Fig. \ref{performance_time2}. We set the FNR constraint $\alpha=0.15$ and the capacity constraint $C=1$.

As illustrated in the left panel of Fig. \ref{performance_time}, all methods converge to the target FNR of $\alpha = 0.15$. However, as seen in the middle panel, CD-CRC and U-CD-CRC eventually converge to satisfy the long-term capacity constraint $C=1$, while D-CRC stabilizes at a value exceeding it, as it is not designed to control the communication overhead. In terms of the objective of problem \eqref{eq_obj}, CD-CRC and D-CRC achieve similar values for the FPR of around $0.4$, while U-CD-CRC yields a higher FPR around $0.6$. This demonstrates the benefits of the adaptive allocation rule \eqref{eq_cap_alloc}.

To gain further insights into the operation of CD-CRC and D-CRC, the top-left corner of Fig. \ref{performance_time2} shows the evolution of the weights $\left\{\beta_k^t\right\}_{k=1}^K$ over time $t$. {\color{blue}It is observed that all schemes quickly identify sensor $1$ as the best due to the online exponentiated gradient rules in \eqref{eq_wei_up_DCRC} and \eqref{eq_wei_up}, which rapidly assign full weight to the sensor due to its superior FPR performance. This way, CD-CRC allocates all communication resources to sensor $1$ via the adaptive resource allocation strategy \eqref{eq_cap_alloc} which allocates communication resources proportionally to the sensors' weights, achieving local performance similar to D-CRC, which does not account for capacity constraint.} CD-CRC meets the capacity constraint by disabling transmissions from non-optimal sensors, while, as shown in the bottom row of Fig. \ref{performance_time2}, D-CRC applies a common local threshold across all sensors, exceeding the capacity constraint. 

As shown in the top-right corner of Fig. \ref{performance_time2}, CD-CRC leverages the update of the global threshold to ensure both FNR and capacity constraints even under stringent capacity constraints (i.e., $C< 2.5$ in Fig. \ref{longPer_C}). U-CD-CRC applies an even communication resource allocation strategy \eqref{eq_even_alloc}, lowering the local thresholds of all sensors to meet the capacity constraint (bottom row of Fig. \ref{performance_time2}), which in turn requires an increase in the global threshold to achieve the target FNR (top-right corner of Fig. \ref{performance_time2}). This results in a higher local FPR for the optimal sensor (sensor $1$) and a corresponding increase in global FPR (see Fig. \ref{performance_time}). While this approach satisfies the capacity constraint, it compromises the performance of the optimal sensor by unnecessarily preserving the performance of the other sensors. 

We now turn to evaluating the performance of CD-CRC, U-CD-CRC, and D-CRC after $T = 913$ time steps as a function of the FNR constraint $\alpha$ given a fixed capacity constraint $C=1.5$. As shown in Fig. \ref{longPer_alpha}, all methods satisfy the FNR constraints for all target values $\alpha$. However, CD-CRC and U-CD-CRC meet the capacity constraint $C$, whereas D-CRC, which does not account for communication limitations, fails to do so. In this regard, note that, as $\alpha$ increases, more pixels need to be recognized as background, leading to an increase in communication load for all three schemes. Furthermore, CD-CRC and D-CRC achieve similar FPR across different $\alpha$ values, while U-CD-CRC, which allocates communication resources equally to all sensors, cannot capitalize on the performance of the optimal sensor, resulting in a higher FPR.

Based on the analysis in Fig. \ref{performance_time2} and Fig. \ref{longPer_alpha}, CD-CRC can achieve the same ideal performance of D-CRC, while also satisfying the capacity constraints by efficiently utilizing the available communication resources. In fact, unlike U-CD-CRC, which evenly distributes capacity, and D-CRC, which ignores capacity constraints, CD-CRC adaptively allocates capacity across the sensors, avoiding unnecessary transmissions and adhering to the capacity constraint.

The impact of the capacity constraint $C$ with a fixed FNR constraint $\alpha=0.2$ is studied in Fig. \ref{longPer_C}. While CD-CRC and U-CD-CRC satisfy the capacity constraint for all values of $C$, D-CRC fails to do so when the capacity constraint is stringent (i.e., $C<2.5$). As compared with U-CD-CRC, CD-CRC with an adaptive capacity allocation requires a much smaller communication load by allocating all resources to the optimal sensor (sensor $1$), while disabling transmission from other non-optimal sensors. Moreover, when $C$ is insufficient even for a single sensor's transmission (i.e., $C<1$), CD-CRC sacrifices the FPR of the optimal sensor to meet the capacity constraint. Conversely, when $C$ is large enough to support transmissions from all sensors, U-CD-CRC achieves the same FPR as CD-CRC and D-CRC.

\section{Conclusions}\label{sec_conclusion}
This paper has introduced communication-constrained distributed conformal risk control (CD-CRC), a novel protocol designed to address multi-label classification problems under both communication constraints and performance guarantees in distributed sensor networks. 

CD-CRC builds on online conformal risk control (CRC) to dynamically adjust both local and global thresholds, ensuring that both false negative rate (FNR) and capacity constraints are satisfied. By integrating online exponentiated gradient optimization, CD-CRC efficiently identifies and prioritizes the most reliable sensors, optimally allocating weights to minimize the false positive rate (FPR). Unlike conventional sensor fusion methods, which optimize average performance criteria under statistical models for target and observations, CD-CRC provides deterministic worst-case performance guarantees, ensuring control over FNR, communication load, and FPR.

The theoretical analysis in this work offers insights into FNR and FPR worst-case guarantees, while simulation results demonstrate that CD-CRC outperforms existing methods, particularly in communication-constrained environments. Overall, CD-CRC presents a promising solution for reliable decision-making in distributed sensor networks under communication constraints, paving the way for further research in this area. Additionally, the CD-CRC framework can be seamlessly integrated with online model updates \cite{orabona2019modern, shalev2012online}, allowing for enhanced adaptability and performance in applications requiring real-time model adjustments. We leave a full investigation of this aspect to future work.

Future work could explore the application of CD-CRC to more complex tasks such as binary segmentation with multiple objects or time-varying target processes with memory. Additionally, accounting for wireless impairments and investigating privacy-preserving techniques, such as differential privacy, would be valuable directions for further study.

\bibliographystyle{IEEEtran}
\bibliography{cite.bib}

\begin{thebibliography}{10}
\providecommand{\url}[1]{#1}
\csname url@samestyle\endcsname
\providecommand{\newblock}{\relax}
\providecommand{\bibinfo}[2]{#2}
\providecommand{\BIBentrySTDinterwordspacing}{\spaceskip=0pt\relax}
\providecommand{\BIBentryALTinterwordstretchfactor}{4}
\providecommand{\BIBentryALTinterwordspacing}{\spaceskip=\fontdimen2\font plus
\BIBentryALTinterwordstretchfactor\fontdimen3\font minus \fontdimen4\font\relax}
\providecommand{\BIBforeignlanguage}[2]{{%
\expandafter\ifx\csname l@#1\endcsname\relax
\typeout{** WARNING: IEEEtran.bst: No hyphenation pattern has been}%
\typeout{** loaded for the language `#1'. Using the pattern for}%
\typeout{** the default language instead.}%
\else
\language=\csname l@#1\endcsname
\fi
#2}}
\providecommand{\BIBdecl}{\relax}
\BIBdecl

\bibitem{behera2019sep}
T.~M. Behera, S.~K. Mohapatra, U.~C. Samal, M.~S. Khan, M.~Daneshmand, and A.~H. Gandomi, ``{I-SEP}: An improved routing protocol for heterogeneous {WSN} for {IoT}-based environmental monitoring,'' \emph{IEEE Internet Things J.}, vol.~7, no.~1, pp. 710--717, Jan. 2020.

\bibitem{muhammad2018secure}
K.~Muhammad, R.~Hamza, J.~Ahmad, J.~Lloret, H.~Wang, and S.~W. Baik, ``Secure surveillance framework for {IoT} systems using probabilistic image encryption,'' \emph{IEEE Trans. Ind. Informat.}, vol.~14, no.~8, pp. 3679--3689, Jan. 2018.

\bibitem{sun2019energy}
G.~Sun, Y.~Liu, Z.~Chen, A.~Wang, Y.~Zhang, D.~Tian, and V.~C. Leung, ``Energy efficient collaborative beamforming for reducing sidelobe in wireless sensor networks,'' \emph{IEEE Trans. Mob. Comput.}, vol.~20, no.~3, pp. 965--982, Nov. 2019.

\bibitem{gong2019optimal}
S.~Gong, S.~Ma, C.~Xing, and G.~Yang, ``Optimal beamforming and time allocation for partially wireless powered sensor networks with downlink {SWIPT},'' \emph{IEEE Trans. Signal Process.}, vol.~67, no.~12, pp. 3197--3212, Apr. 2019.

\bibitem{hansen2017driver}
J.~H. Hansen, C.~Busso, Y.~Zheng, and A.~Sathyanarayana, ``Driver modeling for detection and assessment of driver distraction: Examples from the {UTDrive} test bed,'' \emph{IEEE Signal Process. Mag.}, vol.~34, no.~4, pp. 130--142, Jul. 2017.

\bibitem{tedeschini2023cooperative}
B.~C. Tedeschini, M.~Brambilla, L.~Barbieri, G.~Balducci, and M.~Nicoli, ``Cooperative lidar sensing for pedestrian detection: Data association based on message passing neural networks,'' \emph{IEEE Trans. Signal Process.}, vol.~71, pp. 3028--3042, Aug. 2023.

\bibitem{kamtue2024phyot}
K.~Kamtue, J.~M. Moura, O.~Sangpetch, and P.~Garcia, ``{PhyOT}: Physics-informed object tracking in surveillance cameras,'' in \emph{Proc. IEEE Int. Conf. Acoust., Speech, Signal Process.}, Apr. 2024, pp. 7030--7034.

\bibitem{wang2021rodnet}
Y.~Wang, Z.~Jiang, Y.~Li, J.-N. Hwang, G.~Xing, and H.~Liu, ``{RODNet}: A real-time radar object detection network cross-supervised by camera-radar fused object {3D} localization,'' \emph{IEEE J. Sel. Top. Signal Process.}, vol.~15, no.~4, pp. 954--967, Feb. 2021.

\bibitem{mao2021moving}
Z.~Mao, H.~Su, B.~He, and X.~Jing, ``Moving source localization in passive sensor network with location uncertainty,'' \emph{IEEE Signal Process. Lett.}, vol.~28, pp. 823--827, Apr. 2021.

\bibitem{hall1997introduction}
D.~L. Hall and J.~Llinas, ``An introduction to multisensor data fusion,'' \emph{Proc. of the IEEE}, vol.~85, no.~1, pp. 6--23, Aug. 1997.

\bibitem{rammelkamp2020low}
K.~Rammelkamp, S.~Schr{\"o}der, S.~Kubitza, D.~S. Vogt, S.~Frohmann, P.~B. Hansen, U.~B{\"o}ttger, F.~Hanke, and H.-W. H{\"u}bers, ``Low-level {LIBS} and {Raman} data fusion in the context of in situ {Mars} exploration,'' \emph{Journal of Raman Spectroscopy}, vol.~51, no.~9, pp. 1682--1701, Sept. 2020.

\bibitem{wang2021multi}
J.~Wang, Y.~Zeng, S.~Wei, Z.~Wei, Q.~Wu, and Y.~Savaria, ``Multi-sensor track-to-track association and spatial registration algorithm under incomplete measurements,'' \emph{IEEE Trans. Signal Process.}, vol.~69, pp. 3337--3350, May. 2021.

\bibitem{gostar2020centralized}
A.~K. Gostar, T.~Rathnayake, R.~Tennakoon, A.~Bab-Hadiashar, G.~Battistelli, L.~Chisci, and R.~Hoseinnezhad, ``Centralized cooperative sensor fusion for dynamic sensor network with limited field-of-view via labeled {multi-Bernoulli} filter,'' \emph{IEEE Trans. Signal Process.}, vol.~69, pp. 878--891, Dec. 2020.

\bibitem{katenka2007local}
N.~Katenka, E.~Levina, and G.~Michailidis, ``Local vote decision fusion for target detection in wireless sensor networks,'' \emph{IEEE Trans. Signal Process.}, vol.~56, no.~1, pp. 329--338, Dec. 2007.

\bibitem{niu2006fusion}
R.~Niu, B.~Chen, and P.~K. Varshney, ``Fusion of decisions transmitted over rayleigh fading channels in wireless sensor networks,'' \emph{IEEE Trans. Signal Process.}, vol.~54, no.~3, pp. 1018--1027, Mar. 2006.

\bibitem{sharma2020sensor}
A.~Sharma and S.~Chauhan, ``Sensor fusion for distributed detection of mobile intruders in surveillance wireless sensor networks,'' \emph{IEEE Sens. J.}, vol.~20, no.~24, pp. 15\,224--15\,231, Jul. 2020.

\bibitem{ravindran2022camera}
R.~Ravindran, M.~J. Santora, and M.~M. Jamali, ``Camera, {LiDAR}, and radar sensor fusion based on {Bayesian} neural network {(CLR-BNN)},'' \emph{IEEE Sens. J.}, vol.~22, no.~7, pp. 6964--6974, Feb. 2022.

\bibitem{su2015multi}
H.~Su, S.~Maji, E.~Kalogerakis, and E.~Learned-Miller, ``Multi-view convolutional neural networks for {3D} shape recognition,'' in \emph{Proc. IEEE Int. Conf. Comput. Vis.}, Dec. 2015, pp. 945--953.

\bibitem{liu2024semantic}
Z.~Liu, C.~Guo, and M.~Zhu, ``Semantic redundancy-aware multi-view edge inference based on rate-distortion in {IoT} systems,'' in \emph{IEEE Wireless Communications and Networking Conference (WCNC)}, Apr. 2024, pp. 1--6.

\bibitem{lan2022progressive}
Q.~Lan, Q.~Zeng, P.~Popovski, D.~G{\"u}nd{\"u}z, and K.~Huang, ``Progressive feature transmission for split classification at the wireless edge,'' \emph{IEEE Trans. Wireless Commun.}, vol.~22, no.~6, pp. 3837--3852, Nov. 2022.

\bibitem{gasparin2024conformal}
M.~Gasparin and A.~Ramdas, ``Conformal online model aggregation,'' \emph{arXiv preprint arXiv:2403.15527}, May. 2024.

\bibitem{gibbs2021adaptive}
I.~Gibbs and E.~Candes, ``Adaptive conformal inference under distribution shift,'' \emph{Proc. Adv. Neural Inf. Process. Syst.}, vol.~34, pp. 1660--1672, Dec. 2021.

\bibitem{de2014follow}
S.~De~Rooij, T.~Van~Erven, P.~D. Gr{\"u}nwald, and W.~M. Koolen, ``Follow the leader if you can, hedge if you must,'' \emph{J. Mach. Learn. Res.}, vol.~15, no.~1, pp. 1281--1316, Apr. 2014.

\bibitem{vovk2005algorithmic}
V.~Vovk, A.~Gammerman, and G.~Shafer, \emph{Algorithmic learning in a random world}.\hskip 1em plus 0.5em minus 0.4em\relax Springer, 2005, vol.~29.

\bibitem{barber2021predictive}
R.~F. Barber, E.~J. Candes, A.~Ramdas, and R.~J. Tibshirani, ``Predictive inference with the jackknife+,'' \emph{Ann. Statist.}, vol.~49, no.~1, pp. 486--507, Feb. 2021.

\bibitem{FedCPQQ}
P.~Humbert, B.~Le~Bars, A.~Bellet, and S.~Arlot, ``One-shot federated conformal prediction,'' in \emph{Proc. Int. Conf. Mach. Learn.}, Jul. 2023, pp. 14\,153--14\,177.

\bibitem{zhu2024federated}
M.~Zhu, M.~Zecchin, S.~Park, C.~Guo, C.~Feng, and O.~Simeone, ``Federated inference with reliable uncertainty quantification over wireless channels via conformal prediction,'' \emph{IEEE Trans. Signal Process.}, Jan. 2024.

\bibitem{bhatnagar2023improved}
A.~Bhatnagar, H.~Wang, C.~Xiong, and Y.~Bai, ``Improved online conformal prediction via strongly adaptive online learning,'' in \emph{Proc. Int. Conf. Mach. Learn.}, Jul. 2023, pp. 2337--2363.

\bibitem{angelopoulos2024online}
A.~N. Angelopoulos, R.~F. Barber, and S.~Bates, ``Online conformal prediction with decaying step sizes,'' \emph{arXiv preprint arXiv:2402.01139}, May. 2024.

\bibitem{gibbs2024conformal}
I.~Gibbs and E.~J. Cand{\`e}s, ``Conformal inference for online prediction with arbitrary distribution shifts,'' \emph{J. Mach. Learn. Res.}, vol.~25, no. 162, pp. 1--36, May. 2024.

\bibitem{angelopoulos2022conformal}
A.~N. Angelopoulos, S.~Bates, A.~Fisch, L.~Lei, and T.~Schuster, ``Conformal risk control,'' \emph{arXiv preprint arXiv:2208.02814}, Apr. 2023.

\bibitem{cohen2024cross}
K.~M. Cohen, S.~Park, O.~Simeone, and S.~S. Shitz, ``Cross-validation conformal risk control,'' in \emph{Proc. IEEE Int. Symp. Inf. Theory. (ISIT)}, Jul. 2024, pp. 250--255.

\bibitem{zecchin2025generalization}
M.~Zecchin, F.~Hellstr{\"o}m, S.~Park, S.~Shamai, and O.~Simeone, ``Generalization and informativeness of weighted conformal risk control under covariate shift,'' \emph{arXiv preprint arXiv:2501.11413}, Jan. 2025.

\bibitem{feldman2022achieving}
S.~Feldman, L.~Ringel, S.~Bates, and Y.~Romano, ``Achieving risk control in online learning settings,'' \emph{arXiv preprint arXiv:2205.09095}, Jan. 2023.

\bibitem{angelopoulos2024conformal}
A.~Angelopoulos, E.~Candes, and R.~J. Tibshirani, ``Conformal {PID} control for time series prediction,'' \emph{Proc. Adv. Neural Inf. Process. Syst.}, vol.~36, Dec. 2023.

\bibitem{feng2021uav}
T.~Feng, L.~Xie, J.~Yao, and J.~Xu, ``{UAV}-enabled data collection for wireless sensor networks with distributed beamforming,'' \emph{IEEE Trans. Wireless Commun.}, vol.~21, no.~2, pp. 1347--1361, Aug. 2021.

\bibitem{rezaee2024survey}
K.~Rezaee, S.~M. Rezakhani, M.~R. Khosravi, and M.~K. Moghimi, ``A survey on deep learning-based real-time crowd anomaly detection for secure distributed video surveillance,'' \emph{Personal and Ubiquitous Computing}, vol.~28, no.~1, pp. 135--151, 2024.

\bibitem{gasparin2024merging}
M.~Gasparin and A.~Ramdas, ``Merging uncertainty sets via majority vote,'' \emph{arXiv preprint arXiv:2401.09379}, Mar. 2024.

\bibitem{orabona2019modern}
F.~Orabona, ``A modern introduction to online learning,'' \emph{arXiv preprint arXiv:1912.13213}, May. 2023.

\bibitem{vovk2001competitive}
V.~Vovk, ``Competitive on-line statistics,'' \emph{International Statistical Review}, vol.~69, no.~2, pp. 213--248, Aug. 2001.

\bibitem{cesa2006prediction}
N.~Cesa-Bianchi and G.~Lugosi, \emph{Prediction, learning, and games}.\hskip 1em plus 0.5em minus 0.4em\relax Cambridge university press, Mar. 2006.

\bibitem{everingham2010pascal}
M.~Everingham, L.~Van~Gool, C.~K. Williams, J.~Winn, and A.~Zisserman, ``The {PASCAL} visual object classes {(VOC)} challenge,'' \emph{International journal of computer vision}, vol.~88, pp. 303--338, 2010.

\bibitem{long2015fully}
J.~Long, E.~Shelhamer, and T.~Darrell, ``Fully convolutional networks for semantic segmentation,'' in \emph{Proc. IEEE Conf. Comput. Vis. Pattern Recog. (CVPR)}, Jun. 2015, pp. 3431--3440.

\bibitem{he2016deep}
K.~He, X.~Zhang, S.~Ren, and J.~Sun, ``Deep residual learning for image recognition,'' in \emph{Proc. IEEE Conf. Comput. Vis. Pattern Recog. (CVPR)}, Jun. 2016, pp. 770--778.

\bibitem{kontoyiannis2013optimal}
I.~Kontoyiannis and S.~Verd{\'u}, ``Optimal lossless data compression: Non-asymptotics and asymptotics,'' \emph{IEEE Trans. Inf. Theory.}, vol.~60, no.~2, pp. 777--795, Nov. 2013.

\bibitem{shalev2012online}
S.~Shalev-Shwartz \emph{et~al.}, ``Online learning and online convex optimization,'' \emph{Foundations and Trends{\textregistered} in Machine Learning}, vol.~4, no.~2, pp. 107--194, 2012.

\end{thebibliography}

\appendix
\subsection{Proof of Positivity for the Corrected Global Threshold in CD-CRC} \label{apdx_range_glo_thre}
We begin by establishing a lower bound for the corrected global threshold $\tilde{\theta}^t$ in \eqref{eq_update_glo} following the Appendix A.1 in \cite{feldman2022achieving}. Suppose, for contradiction, that $\tilde{\theta}^t<\delta-\mu(1-\alpha)$ at some time $t$, while for all $t'<t$, we have $\tilde{\theta}^{t'}\geq\delta-\mu(1-\alpha)$. According to the update rule \eqref{eq_update_glo}, we have
\begin{align}
    \tilde{\theta}^{t-1} = \tilde{\theta}^t + \mu\left(N^{t-1}-\alpha\right)\leq\tilde{\theta}^t+\mu\left(1-\alpha\right)<\delta.
\end{align}
Given that $N^{t-1}=0\leq\alpha$ when $\tilde{\theta}^{t-1}<\delta$ ($\theta^{t-1}<0$), we have
\begin{align}
    \tilde{\theta}^t = \tilde{\theta}^{t-1} - \mu\left(N^{t-1}-\alpha\right)\geq \tilde{\theta}^{t-1}\geq \delta-\mu(1-\alpha),
\end{align}
which contradicts the initial assumption. Thus, $\tilde{\theta}^t$ must satisfy
\begin{align}\label{eq_lb_tilde_glo_thre}
    \tilde{\theta}^t\geq\delta-\mu(1-\alpha)>0,
\end{align}
where the last inequality uses the setting \eqref{eq_lb_hyp_glo_pre}.

\subsection{Proof of Theorem \ref{Theo_FNR}} \label{apdx_FNR_gua}
By substituting $t=1,2,\ldots,T$ into the update rule in \eqref{eq_update_glo} and by summing up the contributions over time $t$, the average global FNR over time $T$ can be upper bounded as
\begin{align}
    \frac{1}{T}\sum_{t=1}^TN^t &= \alpha+\frac{\tilde{\theta}^1-\tilde{\theta}^{T+1}}{\mu T}\nonumber\\
    &\leq\alpha + \frac{\tilde{\theta}^1-\delta+\mu\left(1-\alpha\right)}{\mu T} < \alpha +\frac{\tilde{\theta}^1}{T},
\end{align}
where the inequalities use \eqref{eq_lb_tilde_glo_thre}.

\subsection{Proof of Theorem \ref{Theo_cap}}\label{apdx_loc_LZ}
Following Appendix A.1 in \cite{feldman2022achieving}, we start by providing an upper bound of the local threshold $\lambda^t_k$ in the following lemma.
\begin{lemma}\label{lemma_lb_loc_thre}
    For any time $t$ and sensor $k$, the local threshold satisfies the inequality $\lambda_k^t\geq-2\gamma$, where $\gamma$ denotes the step size in the update rule \eqref{eq_loc_thre_up}.
\end{lemma}
\textit{Proof:}
Assume, contradicting the desired inequality, that there exists a time $t$ such that $\lambda_k^t<-2\gamma$. Further assume that for all $t'<t$, we have $\lambda^{t'}_k\geq -2\gamma$. According to \eqref{eq_loc_thre_up} and given that $B^{t-1}_k\in[0,1]$, $C^{t-1}_k\in[0,C]$, $N^{t-1}\in[0,1]$, and $\alpha\in[0,1]$, we have
\begin{align}
    \lambda_k^{t-1} = \lambda^t_k + \gamma\Delta\lambda^{t-1}_k\leq \lambda^t_k +\gamma < -\gamma< 0.
\end{align}

Based on the fact that $B^{t-1}_k=0\leq C^{t-1}_k$ when $\lambda_k^{t-1}<0$, we further consider the following two situations.

If $\lambda_k^{t-1}\geq-\gamma$, we have
\begin{align}
    \lambda^t_k &= \lambda_k^{t-1}-\gamma\max\left\{N^{t-1}-\alpha,B^{t-1}_k-C^{t-1}_k\right\}\nonumber\\
    &\geq \lambda^{t-1}_k-\gamma\geq-2\gamma,
\end{align}
which contradicts our initial assumption.

If $\lambda_k^{t-1}<-\gamma$, we have
\begin{align}
    \lambda^t_k &= \lambda_k^{t-1}\geq-2\gamma,
\end{align}
which again contradicts our initial assumption.

We now prove Theorem \ref{Theo_cap}. According to \eqref{eq_loc_thre_up}, we have the inequality
\begin{align}\label{eq_Delta_lb}
    \Delta\lambda^t_k\geq B^t_k-C^t_k.
\end{align}
By substituting $t=1,2,\ldots,T$ and $k=1,2\ldots,K$ into the update rule in \eqref{eq_loc_thre_up} and by summing up all contributions over time $t$, the average overall required communication costs for all sensors over time $T$ can be expressed as
\begin{align}
    \frac{1}{T}\sum_{t=1}^T\sum_{k=1}^K B^t_k&\leq C+\sum_{t=1}^T\sum_{k=1}^K\Delta\lambda^t_k\nonumber\\
    &=C+\frac{\sum_{k=1}^K\left(\lambda^1_k-\lambda^{T+1}_k\right)}{\gamma T}\nonumber\\
    &\leq C+\frac{\sum_{k=1}^K\left(\lambda^1_k+2\gamma\right)}{\gamma T},
\end{align}
where the first inequality uses \eqref{eq_Delta_lb} and the third inequality uses Lemma \ref{lemma_lb_loc_thre}.

\subsection{Proof of Lemma \ref{lemma_up_FPR}} \label{apdx_FPR_ub}
 This proof generalizes the proof of Theorem 2.14 in \cite{gasparin2024conformal}. By substituting \eqref{eq_glo_prob_vec} and \eqref{eq_glo_thre_offset} into \eqref{eq_glo_pred_vec}, the $l$-th entry in $V^t$ can be expressed as
\begin{align}
    V^t_l = \mathds{1}\left\{\sum_{k=1}^K\beta_k^tU_{k,l}^t+\delta\geq\tilde{\theta}^t\right\}.
\end{align}
The global FPR in \eqref{eq_FPR} can be upper bounded as
\begin{align}
    P^t &= \frac{\sum_{l=1}^L\left(1-Y^t_l\right)\mathds{1}\left\{\sum_{k=1}^K\beta_k^tU_{k,l}^t+\delta\geq\tilde{\theta}^t\right\}}{\sum_{l=1}^L\left(1-Y^t_l\right)}\nonumber\\
    &=\frac{\sum_{l=1}^L\left(1-Y^t_l\right)\mathds{1}\left\{1/\tilde{\theta}^t\sum_{k=1}^K\beta_k^tU_{k,l}^t+\delta/\tilde{\theta}^t\geq 1\right\}}{\sum_{l=1}^L\left(1-Y^t_l\right)}\nonumber\\
    &\leq\frac{1}{\tilde{\theta}^t}\frac{\sum_{l=1}^L\left(1-Y^t\right)\sum_{k=1}^K\beta_k^tU_{k,l}^t}{\sum_{l=1}^L\left(1-Y^t_l\right)}+\frac{\delta}{\tilde{\theta}^t}\nonumber\\
    &=\frac{1}{\tilde{\theta}^t}\sum_{k=1}^K\beta^t_k\frac{\sum_{l=1}^L\left(1-Y^t\right)U^t_{k,l}}{\sum_{l=1}^L\left(1-Y^t_l\right)} + \frac{\delta}{\tilde{\theta}^t}\nonumber\\
    &=\frac{1}{\tilde{\theta}^t}\sum_{k=1}^K\beta^t_kP^t_k + \frac{\delta}{\tilde{\theta}^t},
\end{align}
where the third inequality uses the fact that $\mathds{1}\left\{x\geq 1\right\}\leq x$ for any $x\geq 0$.

\subsection{Proof of Theorem \ref{Theo_FPR}} \label{apdx_FPR_ub2}

We begin by determining the range of the corrected global threshold $\tilde{\theta}^t$ in CD-CRC. By applying a contradiction method similar to that used in Appendix \ref{apdx_range_glo_thre}, we can establish the following bounds,
\begin{align}\label{eq_range_glo_thre}
    \delta-\mu(1-\alpha)\leq\tilde{\theta}^t\leq 1+\delta+\mu\alpha.
\end{align}

As a result, the average global FPR is upper bounded as
\begin{align}\label{eq_apdx_upbpund2}
    &\frac{1}{T}\sum_{t=1}^T P^t\leq\frac{1}{T}\sum_{t=1}^T\sum_{k=1}^K\beta^t_k\frac{P^t_k}{\tilde{\theta}^t} + \frac{1}{T}\sum_{t=1}^T\frac{\delta}{\tilde{\theta}^t}\nonumber\\
    \leq& \frac{2}{T}\hspace{-1mm}\left(\hspace{-1mm}\ln K\cdot\frac{\max\limits_{t\in\left\{1,\ldots,T\right\}}\hspace{-1.5mm}\Delta P^t/\tilde{\theta}^t}{\sum_{t=1}^T\Delta P^t/\tilde{\theta}^t}\right)^{\hspace{-1mm}\frac{1}{2}} \hspace{-2mm}\cdot \hspace{-1mm}\left(\sum_{t=1}^T \frac{P^t_{\textrm{max}}}{\tilde{\theta}^t}-\hspace{-3mm}\min_{k\in\left\{1,\ldots,K\right\}} \hspace{-0.5mm}\sum_{t=1}^T\frac{P_k^t}{\tilde{\theta}^t}\right)^{\hspace{-1mm}\frac{1}{2}}
    \nonumber\\
    &\hspace{-2mm}\cdot\left(\min_{k\in\left\{1,\ldots K\right\}}\sum_{t=1}^T\frac{P_k^t}{\tilde{\theta}^t}-\sum_{t=1}^T\frac{P^t_{\textrm{min}}}{\tilde{\theta}^t}\right)^{\hspace{-1mm}\frac{1}{2}}\nonumber\\
    &\hspace{-2mm}+\frac{1}{T}\hspace{-1mm}\left(\frac{16}{3}\ln K+2\right)\hspace{-1.5mm}\max_{t\in\left\{1,\ldots,T\right\}}\hspace{-2mm}\frac{\Delta P^t}{\tilde{\theta}^t} \hspace{-0.5mm}+\frac{1}{T}\min_k\sum_{t=1}^T\hspace{-0.5mm}\frac{P_k^t}{\tilde{\theta}^t} + \frac{1}{T}\sum_{t=1}^T\frac{\delta}{\tilde{\theta}^t}\nonumber\\
    \leq & \frac{2}{T}\left(\frac{\left(1+\delta+\mu\alpha\right)\ln K}{\delta-\mu(1-\alpha)}\cdot\frac{\max\limits_{t\in\left\{1,\ldots, T\right\}}\Delta P^t}{\sum_{t=1}^T\Delta P^t}\right)^{\hspace{-1mm}\frac{1}{2}}\nonumber\\
    &\hspace{-2mm}\cdot\left(\frac{\sum_{t=1}^TP^t_{\text{\rm{max}}}}{\delta-\mu(1-\alpha)}-\frac{\min\limits_{k\in\left\{1,\ldots,K\right\}}\sum_{t=1}^TP_k^t}{1+\delta+\mu\alpha}\right)^{\hspace{-1mm}\frac{1}{2}}\nonumber\\
    &\hspace{-2mm}\cdot\left(\frac{\min\limits_{k\in\left\{1,\ldots,K\right\}}\sum_{t=1}^TP_k^t}{\delta-\mu(1-\alpha)}-\frac{\sum_{t=1}^TP^t_\text{\rm{min}}}{1+\delta+\mu\alpha}\right)^{\hspace{-1mm}\frac{1}{2}}\nonumber\\
    &\hspace{-2mm}+\frac{1}{T\left(\delta-\mu(1-\alpha)\right)} \left(\frac{16}{3}\ln K+2\right) \hspace{-1mm}\max\limits_{t\in\left\{1,\ldots,T\right\}}\hspace{-0.2cm}\Delta P^t \nonumber\\
    &\hspace{-2mm}+\frac{\mathcal{P}^*}{\delta-\mu(1-\alpha)} + \frac{1}{T}\sum_{t=1}^T\frac{\delta}{\tilde{\theta}^t},
\end{align}
where the first inequality uses Lemma \ref{lemma_up_FPR}, the second inequality extends from Theorem \ref{theo_sim} (see also Theorem 8 in \cite{de2014follow}), and the third inequality uses \eqref{eq_range_glo_thre}.

\subsection{Additional Experiments} \label{apdx_experiment}

\begin{figure*}[t]
    \centering
    {
    \includegraphics[width = 0.325\textwidth]{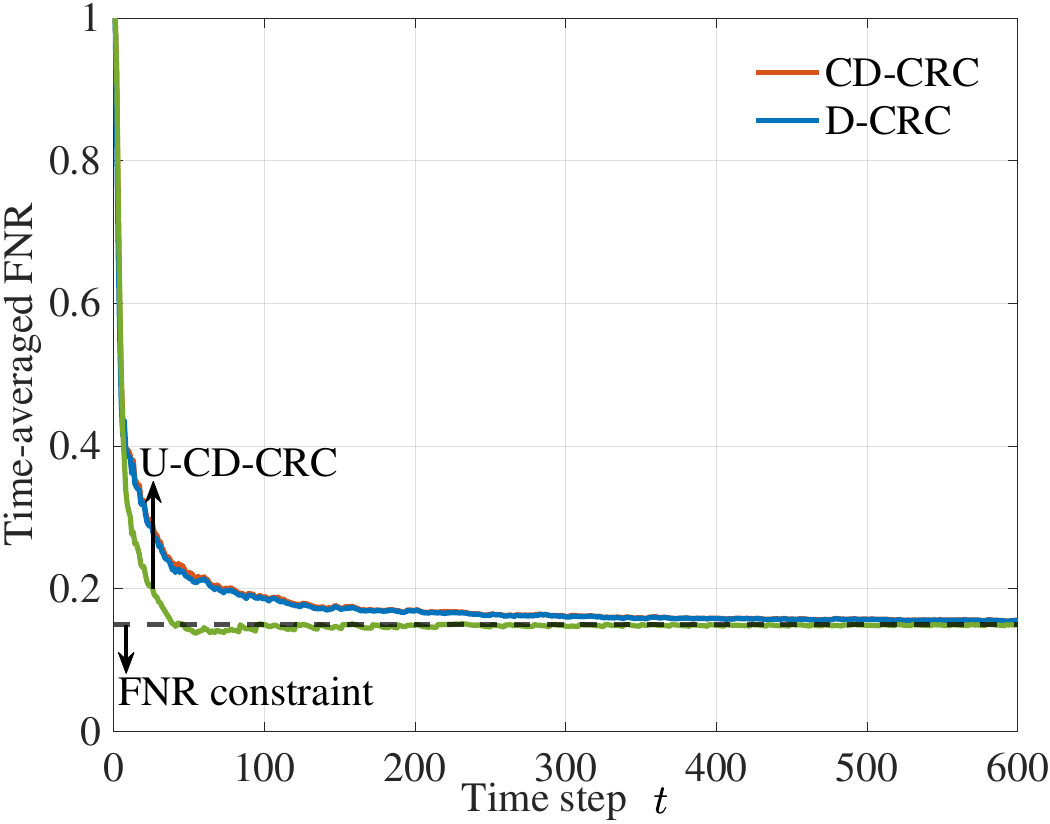}
    \includegraphics[width = 0.325\textwidth]{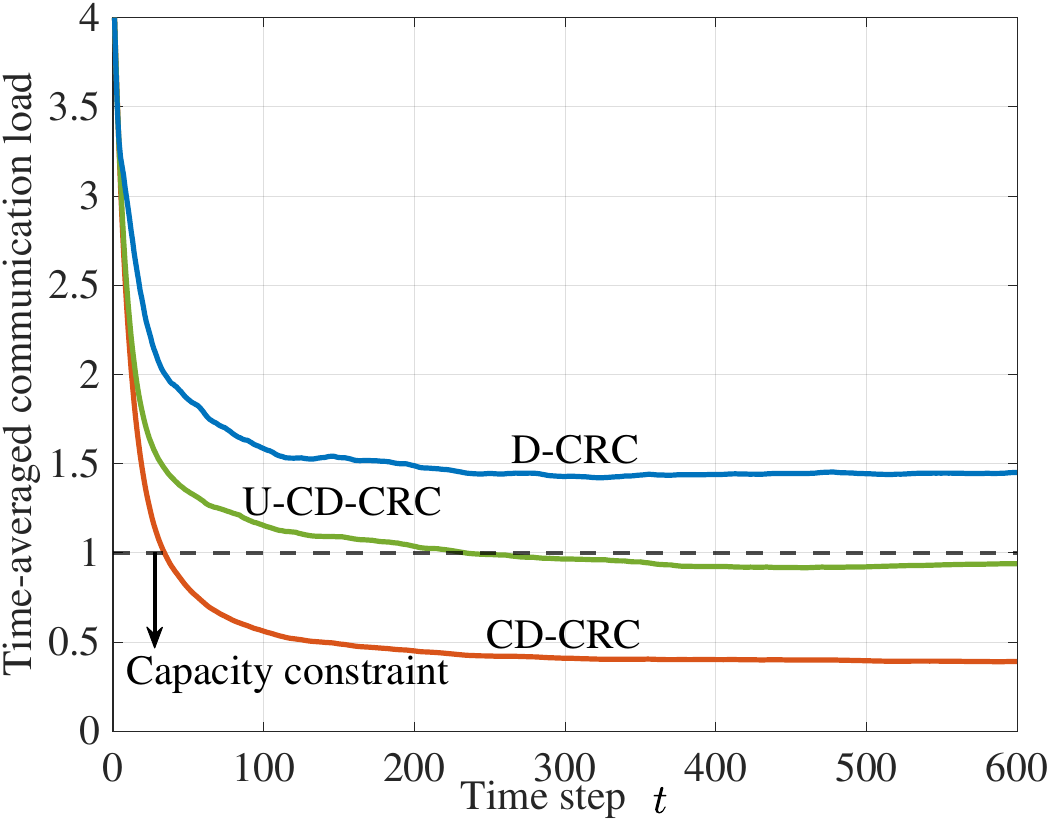}
    \includegraphics[width = 0.325\textwidth]{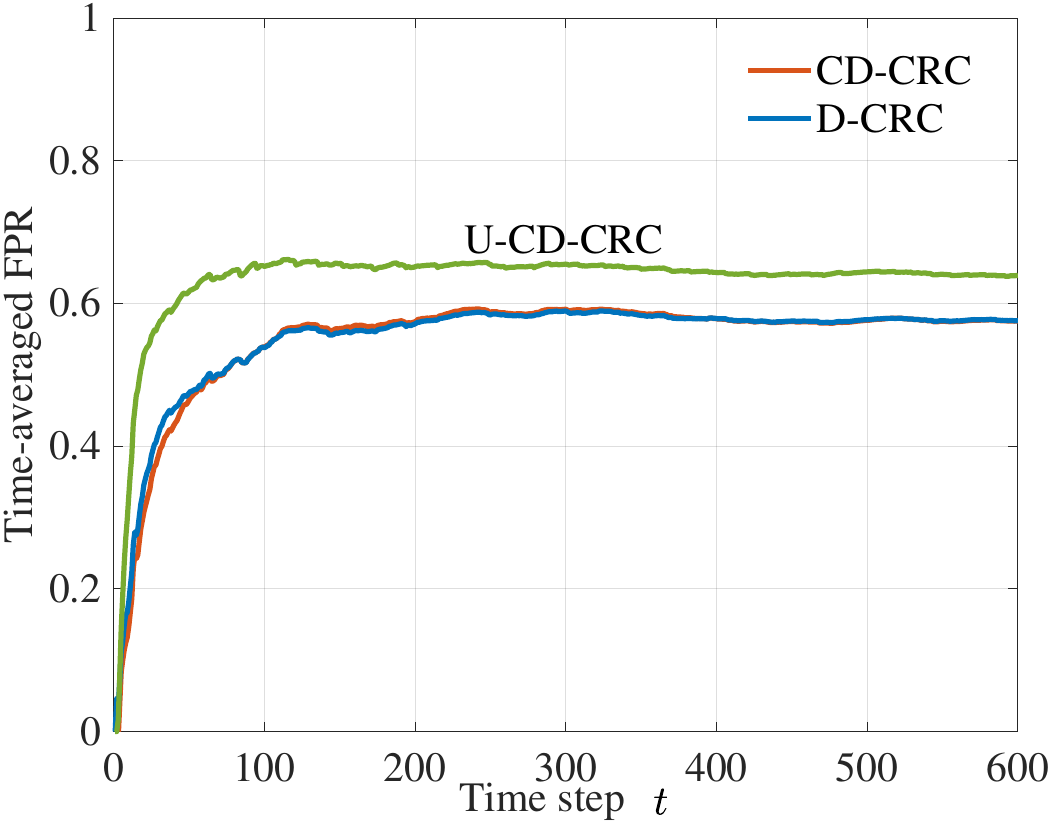}
    }
    \caption{Time-averaged FNR (left), time-averaged communication load (center), and time-averaged FPR (right) for CD-CRC, U-CD-CRC and D-CRC \cite{gasparin2024conformal} over time $t$ with FNR constraint $\alpha=0.15$ and capacity constraint $C=1$.}
    \label{performance_time_apdx1}
\end{figure*}

\begin{figure*}[t]
    \centering
    {
    \includegraphics[width = 0.315\textwidth]{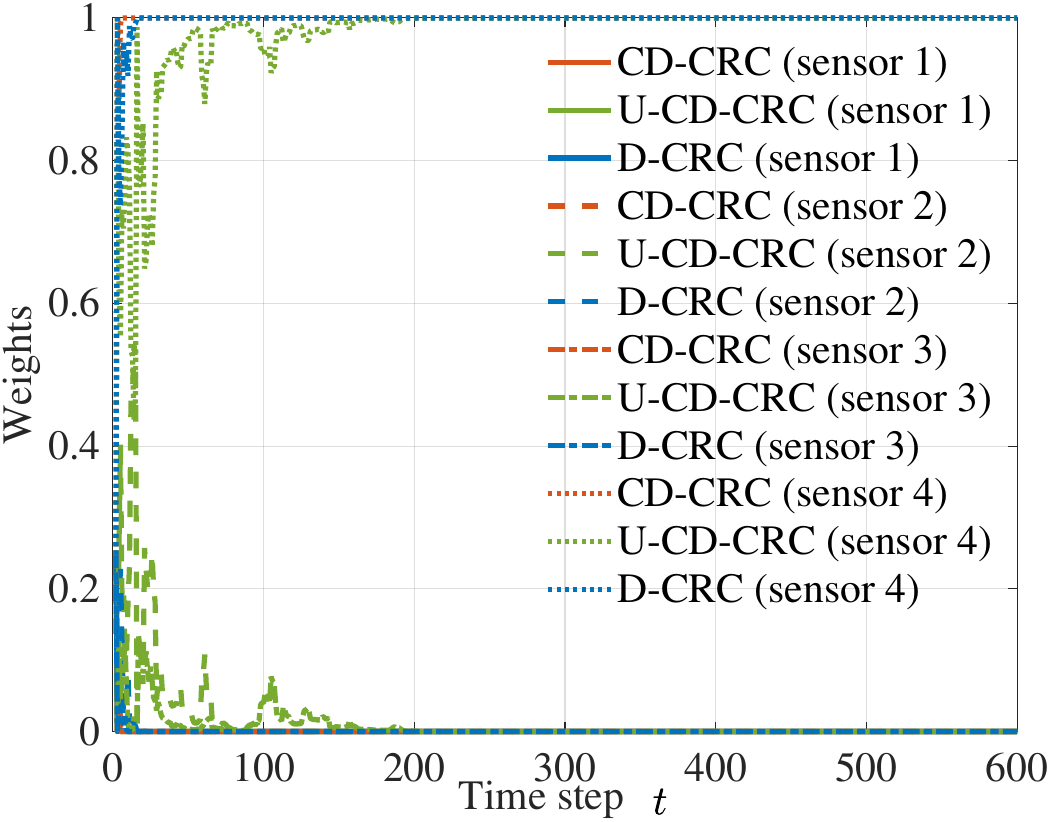}
    \includegraphics[width = 0.32\textwidth]{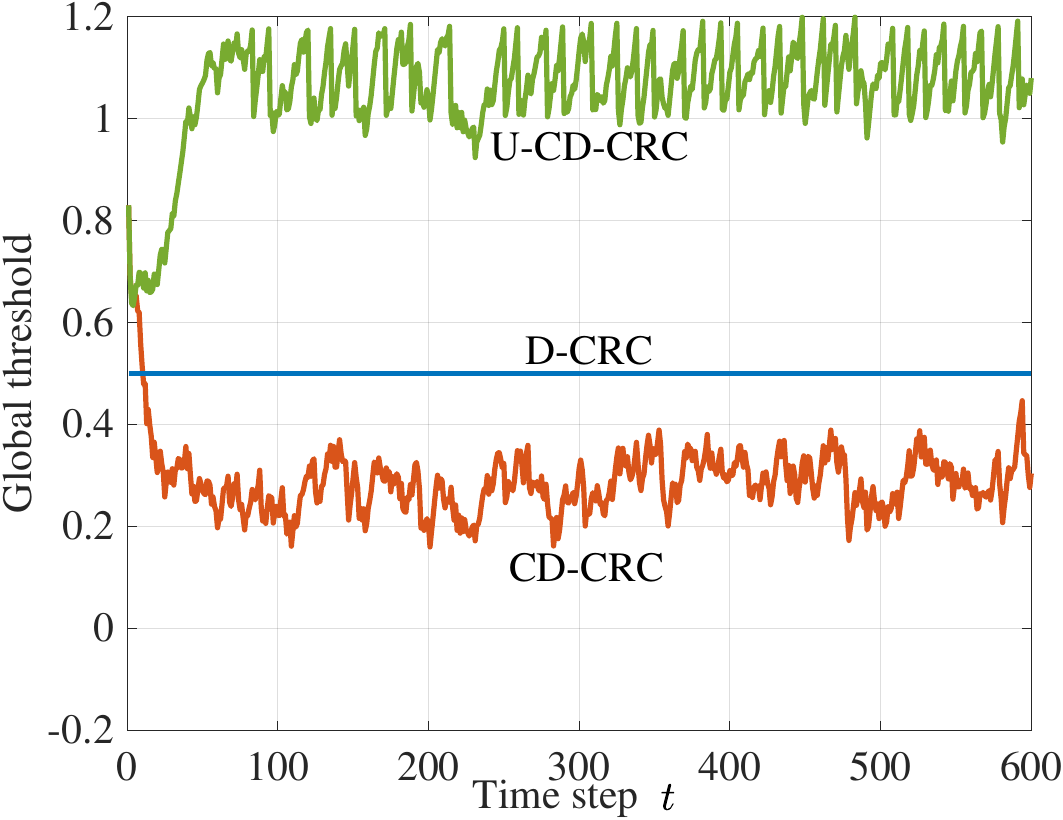}
    \includegraphics[width = 0.32\textwidth]{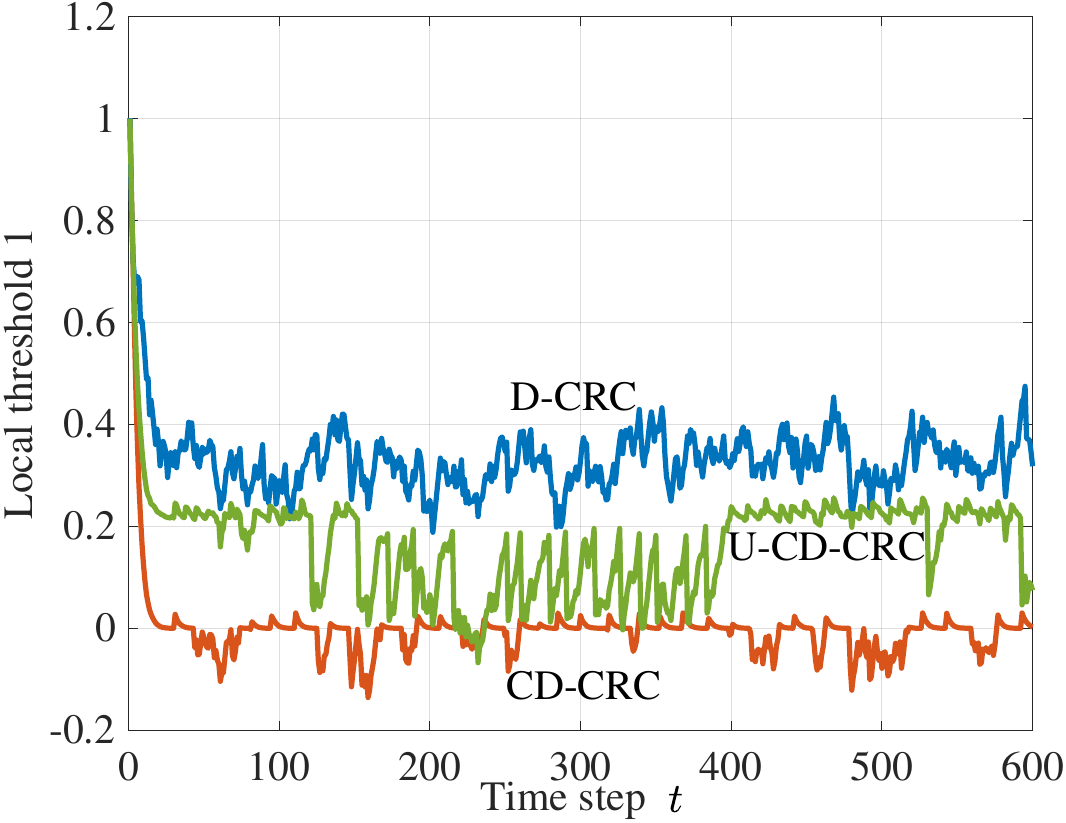}
    \includegraphics[width = 0.32\textwidth]{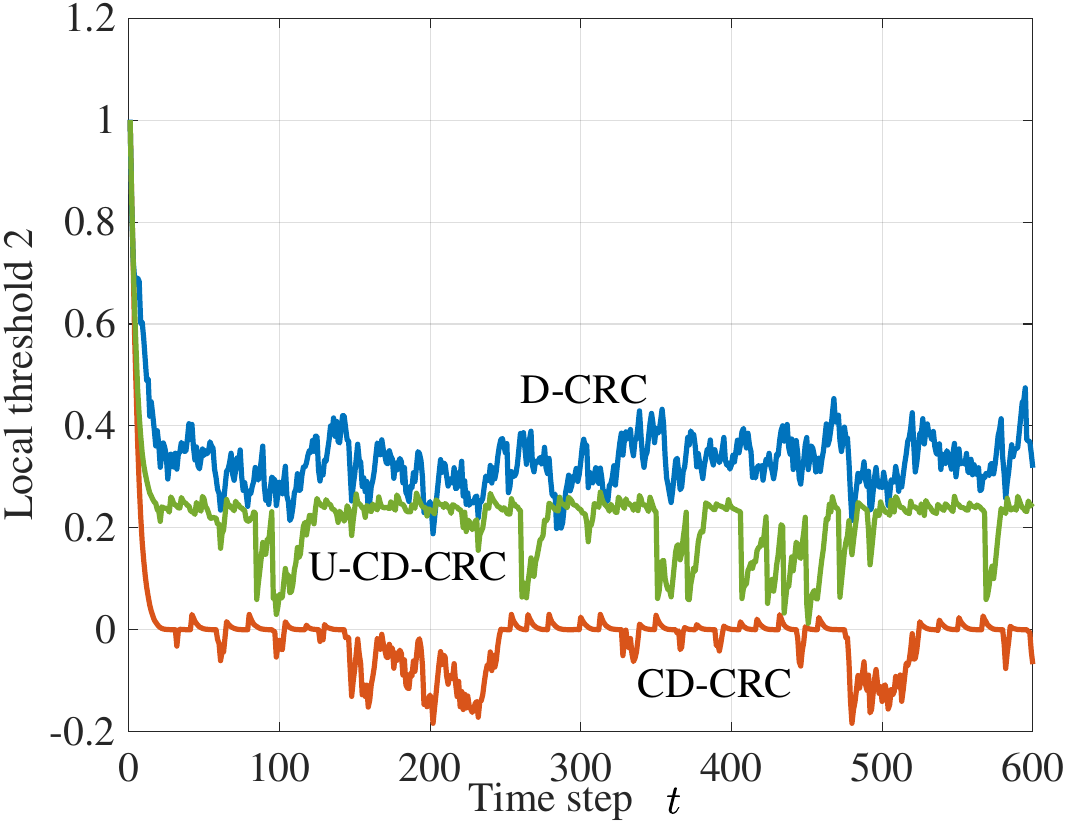}
    \includegraphics[width = 0.32\textwidth]{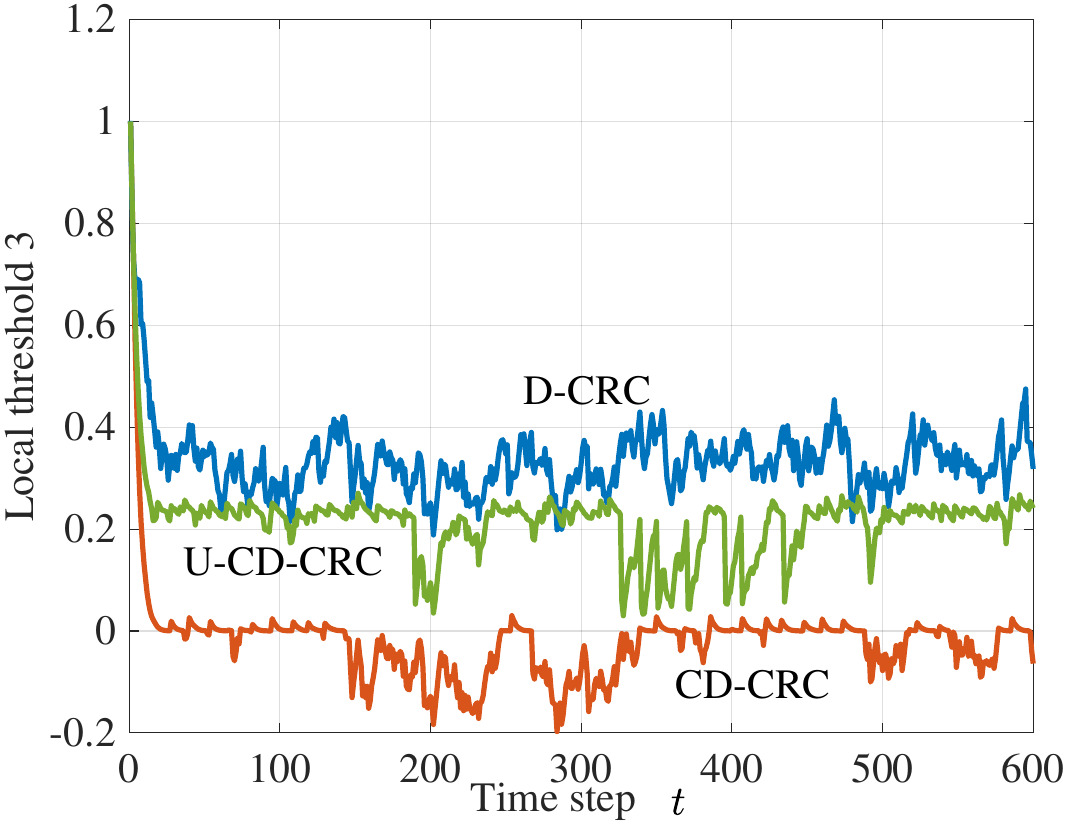}
    \includegraphics[width = 0.32\textwidth]{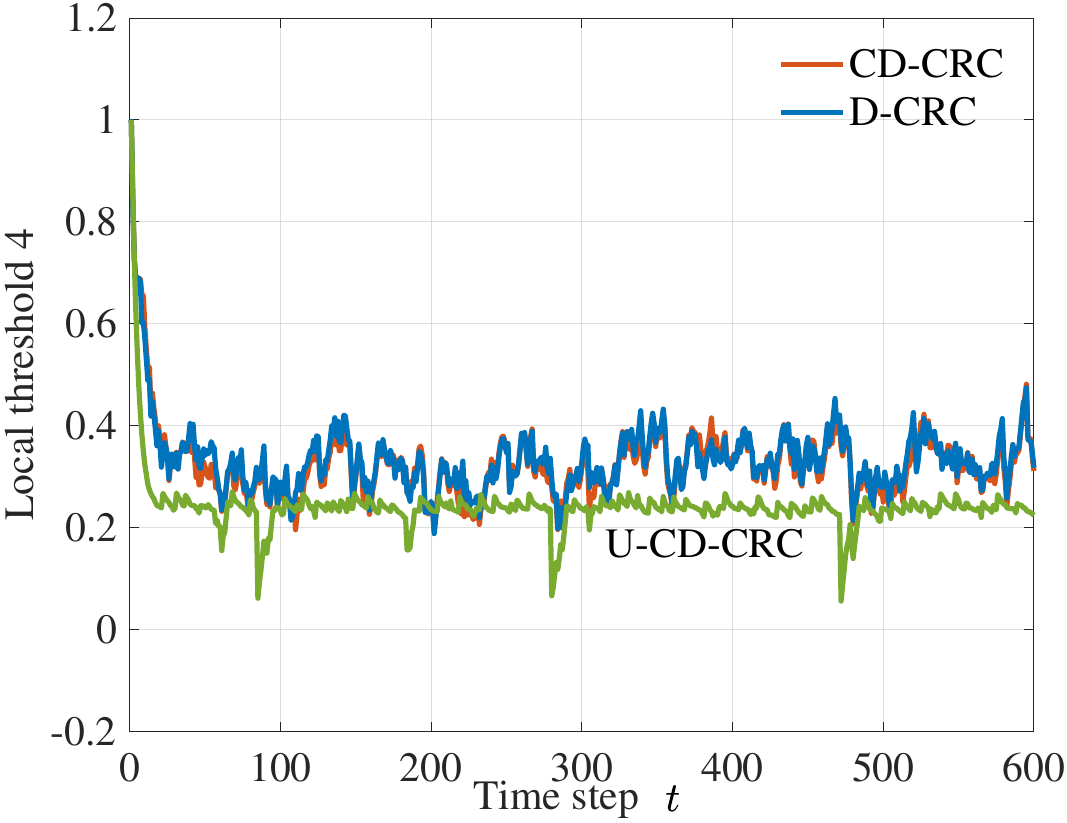}
    }
    \caption{Evolution of the weights (top-left), global thresholds (top-middle), local thresholds for sensor $1$ (top-right), local thresholds for sensor $2$ (bottom-left), local thresholds for sensor $3$ (bottom-middle), and local thresholds for sensor $4$ (bottom-right) for CD-CRC, U-CD-CRC and D-CRC \cite{gasparin2024conformal} over time $t$ with FNR constraint $\alpha=0.15$ and capacity constraint $C=1$.}
    \label{performance_time_apdx2}
\end{figure*}

To provide further insights into the performance of CD-CRC, we conduct additional experiments in which the base sensor (sensor $1$) is set to observe a smaller random $40\times 40$ region, while the other three sensors are still set to observe random $60\times 60$ regions, each subject to different disturbances, including brightness adjustment, noise addition, and blur effects. Note that in this case there is no a priori best sensor. The results are shown in Fig. \ref{performance_time_apdx1}, reporting the time-averaged FNR, the communication load, and the FPR, as well as in Fig. \ref{performance_time_apdx2}, showing the evolution of the local thresholds, the global thresholds, and the weights over time $t = 600$.

From Fig. \ref{performance_time_apdx1}, it is observed that all methods converge to the target FNR of $\alpha = 0.15$. CD-CRC and U-CD-CRC converge to satisfy the long-term capacity constraint $C = 1$, whereas D-CRC exceeds this constraint due to its lack of communication overhead control. In terms of FPR, CD-CRC and D-CRC achieve similar values around $0.6$, while U-CD-CRC yields a higher FPR around $0.65$, demonstrating the benefits of the adaptive allocation rule in equation \eqref{eq_cap_alloc}. These results show similar trends to the original experimental setup studied in Sec. \ref{subsec_performance}.

Fig. \ref{performance_time_apdx2} illustrates that the learned weights for all methods initially fluctuate among sensors $2$, $3$, and $4$, before converging entirely to sensor $4$, indicating that sensor $4$ is identified as the optimal sensor. U-CD-CRC converges more slowly due to the uniform capacity allocation limiting the capacity allotted to the optimal sensor.

Other threshold evolution trends align with those observed in the original experimental setup. In particular, by allocating all the resources to sensor $4$ and disabling the transmission of other non-optimal sensors, CD-CRC achieves comparable FPR performance to the ideal benchmark D-CRC. Unlike D-CRC, however, CD-CRC also satisfies both the capacity constraint and the FNR constraint by simultaneously adjusting the global threshold, whereas U-CD-CRC sacrifices FPR to meet the capacity constraint.

These results demonstrate that, even when there is no a priori optimal sensor, all three methods eventually converge to allocate the full weight to a single sensor. This outcome results from the use of the online exponentiated gradient rule \eqref{eq_wei_up_DCRC}, which aims at identifying quickly and robustly to the optimal expert, i.e., sensor \cite{gasparin2024conformal, de2014follow}.

\end{document}